%% file: main.tex
\documentclass{article}
\usepackage[utf8]{inputenc}

\usepackage{amsmath}
\usepackage[figuresleft]{rotating}
\usepackage{float}
\usepackage{wrapfig}
\usepackage{graphicx}
\usepackage{hyperref}

\usepackage[table]{xcolor}

\usepackage{multirow}
\usepackage{subcaption} 
\usepackage{amsfonts}
\usepackage{geometry}
\usepackage[vlined,ruled,linesnumbered]{algorithm2e}

\usepackage{tikz}
\tikzstyle{market} = [rectangle, rounded corners, minimum width=2cm, minimum height=1cm,text centered, draw=black, fill=blue!30]
\tikzstyle{agent} = [rectangle, minimum width=2cm, minimum height=1cm,text centered, draw=black, fill=red!30]
\tikzstyle{data} = [rectangle, minimum width=2cm, minimum height=1cm,text centered, draw=black, fill=green!30]
\tikzstyle{arrow} = [thick,->,>=stealth]

\usepackage{pgfplots}
\pgfplotsset{every tick label/.append style={font=\footnotesize}}
\pgfplotsset{every y axis label/.append style={font=\footnotesize}}
\pgfplotsset{every x axis label/.append style={font=\footnotesize}}
\newlength\figH
\newlength\figW
\pgfplotsset{compat=1.17}

\usepackage{enumitem}
\setitemize{nolistsep,leftmargin=*}

\title{Traders in a Strange Land: Agent-based discrete-event market simulation of the Figgie card game}
\author{Anthony Ozerov, Steven DiSilvio, Yu\;(Anna) Luo}
\date{August 12, 2021}

\begin{document}

\maketitle
\begin{abstract}
Figgie is a card game that approximates open-outcry commodities trading. We design strategies for Figgie and study their performance and the resulting market behavior. To do this, we develop a flexible agent-based discrete-event market simulation in which agents operating under our strategies can play Figgie. Our simulation builds upon previous work by simulating latencies between agents and the market in a novel and efficient way. The fundamentalist strategy we develop takes advantage of Figgie's unique notion of asset value, and is, on average, the profit-maximizing strategy in all combinations of agent strategies tested. We develop a strategy, the ``bottom-feeder'', which estimates value by observing orders sent by other agents, and find that it limits the success of fundamentalists. We also find that chartist strategies implemented, including one from the literature, fail by going into feedback loops in the small Figgie market. We further develop a bootstrap method for statistically comparing strategies in a zero-sum game. Our results demonstrate the wide-ranging applicability of agent-based discrete-event simulations in studying markets.
\end{abstract}

\section{Introduction}

Until 2019, there were ``no broadly available high-fidelity market simulation environments'' \cite{abides}, preventing research into the interaction of agents like the high-frequency traders of May 6, 2010 Flash Crush. Agent-based simulations like ABIDES \cite{abides} (in which traders interact by placing buy and sell orders on a market) are useful in helping to understand the interactions between traders. They also provide a space to examine strategies in which—unlike when historical data is used—implementing a strategy affects the market.

Existing research on trader interaction has focused on how different proportions of different simplified types of traders affect the behavior of prices \cite{heterogenous}, and how simulating trades between these types of agents can recreate various statistical properties of real markets \cite{bitcoin}\cite{flashcrash_model}. This research has shown the effectiveness of agent-based modeling when applied to markets.

Figgie is a card game that simulates a market with just 4 assets and 4 traders \cite{figgie}. We develop trading strategies for this simple version of a market and study how they perform in simulations, grounding our work in similar simulations of real markets. Expanding on previous work, we examine not just the resulting behavior of the market, but the performance of different strategies under different simulation settings.

We discuss multiple analogues of real trading strategies as applied to Figgie, including fundamentalists, chartists, noise traders, and a social trading strategy based on the completely transparent market of Figgie.

\section{Methods}

\subsection{Simulation}
\paragraph{Framework}
The simulation is an agent-based, discrete-event simulation. Every trader in the market is represented as a separate agent in the simulation, and an event can happen at any time in $\mathbb{R}^+$. There are two kinds of events that can occur:
\begin{itemize}
    \item \textbf{Consideration event}. In this event, an agent takes a look at the market, and either decides to make an order, in which case an Add Order event is added to the queue, or decides to not make an order, in which case a new Consideration event is added to the queue. When a new Consideration event is being created for agent $i$, its time is $t_0 + X$, where $X\sim \text{Exp}(C_i)$ and $E(X)=1/C_i$. $t_0$ is the current simulation time and $C_i$ is the agent's consideration rate.
    \item \textbf{Add Order event}. In this event, a buy/sell order for asset $j$ is added to the corresponding buy/sell queue in asset $j$. Since this is the only time that asset $j$'s order book can change, this is also when asset $j$'s order-matching mechanism functions. Once the Add Order event finishes, a new Consideration event is created for the agent. Thus the assumption is that agents will only be willing to start considering making a trade again once their order has entered the order book.
\end{itemize}
The event-handling behavior is implemented efficiently in the simulation by storing events in a min-heap ordered by event time. This makes the time complexity of finding the next event $\Theta(1)$ and that of adding an event $O(\log n)$, where $n$ is the number of events in the heap \cite{algos}.

\paragraph{Order-matching mechanism}
Every time an Add Order event occurs, the order-matching mechanism for the corresponding asset runs. It fulfills matching buy and sell orders and removes them from the order book.

\begin{algorithm}[H]
\DontPrintSemicolon
 Get the lowest buy order $b$ and the highest sell order $s$, set $b_p$ and $s_p$ to their prices respectively\;
 \While{$b_p\geq s_p$}{
  Set $v$ = min(buy volume, sell volume, seller inventory)\;
  \eIf{$v>0$}{
   Subtract $v$ from the seller's inventory and add it to the buyer's\;
   Subtract $v\times b_p$ from the buyer's cash and add it to the seller\;
  }{
   No trade happens\;}
  \If {b or s has not been fully fulfilled by the trade}{add the unfulfilled order back to the order book with the remaining volume to be fulfilled\;}
 Get the lowest buy order $b$ and the highest sell order $s$, set $b_p$ and $s_p$ to their prices respectively\;
 }
 Add $b$ and $s$ back to the order book\;
 \caption{Order-matching mechanism}
 \label{algo:matching}
\end{algorithm}
\noindent Algorithm \ref{algo:matching} is implemented efficiently in the simulation by storing the buy and sell orders in min-heaps and max-heaps respectively. As in the event handler, this gives $\Theta(1)$ time complexity for finding the next order to check and $O(\log n)$ time complexity for adding an order, where $n$ is the number of orders in the order book \cite{algos}.

\paragraph{Latency} Latency is the difference in time between when the buyer/seller sends order and the order book receives it. The simulation can handle latencies between trader agents and the market. Unlike ABIDES, which implements a latency matrix containing latencies between every pair of agents \cite{abides}, this means that market information does not need to be passed in messages between the exchange and the traders, as seeing time-delayed information is easily simulated.

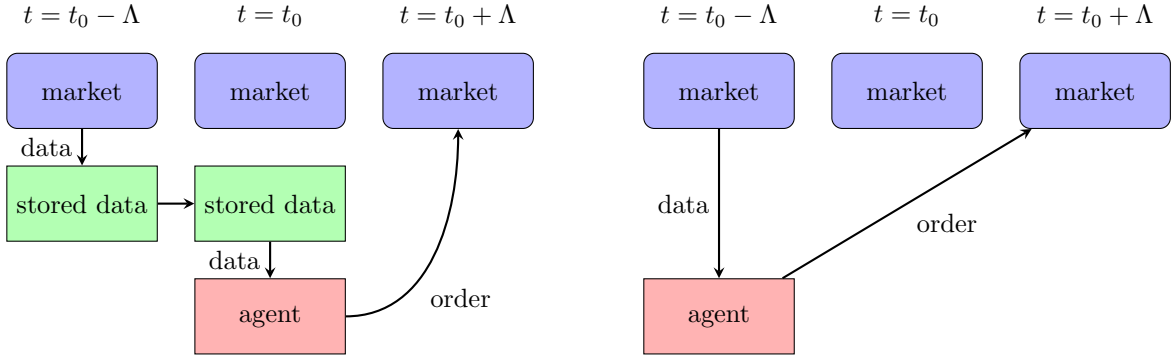
\begin{figure}[!htp]
\centering
\begin{subfigure}{.48\textwidth}
    \centering
    \begin{tikzpicture}
    \node[market] (m1){market};
    \node[market] (m2) [right of=m1, xshift=1.5cm]{market};
    \node[market] (m3) [right of=m2, xshift=1.5cm]{market};
    \node[data] (d1) [below of=m1, yshift=-0.5cm]{stored data};
    \node[data] (d2) [below of=m2, yshift=-0.5cm]{stored data};
    \node (t1) [above of=m1] {$t=t_0-\Lambda$};
    \node (t2) [above of=m2] {$t=t_0$};
    \node (t2) [above of=m3] {$t=t_0+\Lambda$};
    \node [agent] (a2) [below of=d2, yshift=-0.5cm] {agent};
    \draw[arrow] (m1) -- (d1) node[midway,left] {data};
    \draw[arrow] (d2) -- (a2) node[midway,left] {data};
    \draw[arrow] (a2) to  [out=0, in=270] (m3) node[midway,below right, xshift=4.5cm, yshift=-2.5cm] {order};
    \draw[arrow] (d1) -- (d2);
    \end{tikzpicture}
    \caption{Inefficient latency simulation. Market data from\\ $t_0-\Lambda$ must be stored or recreated for use at $t_0$.}
    \label{fig:latencya}
\end{subfigure}%
\hfill\begin{subfigure}{.48\textwidth}
    \centering
    \begin{tikzpicture}
    \node[market] (m1){market};
    \node[market] (m2) [right of=m1, xshift=1.5cm]{market};
    \node[market] (m3) [right of=m2, xshift=1.5cm]{market};
    \node (t1) [above of=m1] {$t=t_0-\Lambda$};
    \node (t2) [above of=m2] {$t=t_0$};
    \node (t2) [above of=m3] {$t=t_0+\Lambda$};
    \node [agent] (a1) [below of=m1, yshift=-2cm] {agent};
    \draw[arrow] (m1) -- (a1) node[midway,left] {data};
    \draw[arrow] (a1) -- (m3) node[midway,below right] {order};
    \end{tikzpicture}
    \caption{Efficient latency simulation. Same outcome as (a), but market data from $t_0-\Lambda$ is accessed directly.}
    \label{fig:latencyb}
\end{subfigure} 
\setlength{\abovecaptionskip}{5pt}
\caption{Comparison of latency simulation methods (a) and (b).}
\label{fig:latency}
\end{figure}

Consider an agent $a$ with a latency of $\Lambda$ which is considering making a trade at time $t_0$. $a$ should see the market as it was at time $t_0-\Lambda$, and, if it decides to add an order, the order will be added to the order book at time $t_0+\Lambda$. There are two ways to implement this:
\begin{enumerate}[label=(\alph*)]
    \item Give $a$ a consideration event at time $t_0$, letting it see historical market data from $t_0-\Lambda$. This would require storing a copy of all of the order books or a data structure that could generate past order books.
    \item Give $a$ a consideration event at time $t_0-\Lambda$, letting it see the market at that time, and delay the add order event by $2\Lambda$ instead of $\Lambda$. In practice, this means that $a$ uses market data from time $t_0-\Lambda$ to come up with an order that is added at time $t_0+\Lambda$, which is exactly the desired result.
\end{enumerate}
These two methods are illustrated in Figure \ref{fig:latency}. (a) was considered but determined to be too inefficient due to the storage of intermediate market data, so (b) is used in the simulation.

\subsection{Figgie Rules and Modifications \cite{figgie} }

Figgie is a trading card game meant to model open-outcry commodities trading.  In practice, either 4 or 5 players can participate in a game, but for the purposes of this paper we only consider the case of 4 player gameplay.  

In Figgie, a card deck consists of 40 cards, with each of the cards being one of 4 suits: spades, clubs, hearts or diamonds.  The composition of the deck, however, varies between games, with the only constraint being that one suit must have 12 cards, another must have 8, and the remaining two must have 10.  There are thus 12 possible decks, which are shown in Table \ref{table:decks}.  Prior to the start of the game, no player has any information on the composition of the deck.

For a given deck composition, the suit consisting of 12 cards is the common suit.  The suit which is of the same color as the common suit, but not the common suit itself, is the goal suit (note that, as in a traditional 52 card deck, hearts and diamonds are both red while spades and clubs are black).  The only cards which give any payoff are those of the goal suit (detailed further below).  Apart from their suit however, individual cards are indistinguishable from each other.

At the start of the game, each player is given 350 chips to make trades with. Each round, each player must ante 50 chips to form the pot (for a 200 chip pot in total) and then a randomly chosen deck is selected, from which each player is dealt 10 cards. Throughout the round, players negotiate prices with each other to trade chips for cards, with the goal of making as much money as possible.  There are 2 ways of making money:
\begin{enumerate}
    \item Profiting off of the trades themselves (sell off your initial cards or resell cards for prices higher than you purchased them)
    \item Collecting goal suit cards
\end{enumerate}

Each round lasts for 4 minutes, and at the end of the round, the goal suit is revealed, and each player receives 10 chips from the pot for each goal suit card which they possess.  Additionally, the player with the majority of the goal suit cards receives the rest of the pot, which works out to 100 chips when the goal suit consists of 10 cards and 120 chips when the goal suit consists of 8 cards.  In the case of ties for majority, the remainder of the pot is instead split.  

In the case of our simulation, we have made some slight modifications to the game:

\begin{itemize}
    \item Agents trade in continuous amounts of cash rather than chips
    \item Rather than lasting 4 minutes, games last 10000 events (detailed in the simulation framework description previously)
    \item Rather than running consecutive rounds where agents have a running total of cash, in each iteration of gameplay an agent starts off with the same amount of cash so that consecutive runs of gameplay can be more easily compared
\end{itemize}

\subsection{Agent Strategies}
In our model, we formulated three different kinds of agents that use different algorithms to perform trades in Figgie. All of the trading strategies implement the same trading algorithm based on a calculated expected value for an asset, as demonstrated in Algorithm 2 below, similar to \cite{heterogenous} but without considering risk, as we assume the goal of the agents is to maximize their expected return.

\newcommand{\commonsuit}{\cellcolor[HTML]{F6F638}}
\newcommand{\goalsuit}{\cellcolor[HTML]{99FF66}}
\begin{table}
\begin{tabular}{ p{1.5cm}||p{1.5cm} p{1.5cm} p{1.5cm} p{1.5cm}| p{1.5cm} p{1.5cm} }
  & Spades & Clubs & Hearts & Diamonds & Majority & Payoff \\
 \hline
 Deck 0   & \commonsuit 12 & \goalsuit 8 & 10 & 10 & 5 & 120 \\
 Deck 1   & \commonsuit 12 & \goalsuit 10 & 8 & 10 & 6 & 100\\
 Deck 2   & \commonsuit 12 & \goalsuit 10 & 10 & 8 & 6 & 100\\
 Deck 3   & \goalsuit 8 & \commonsuit 12 & 10 & 10 & 5 & 120\\
 Deck 4   & \goalsuit 10 & \commonsuit 12 & 8 & 10 & 6 & 100\\
 Deck 5   & \goalsuit 10 & \commonsuit 12 & 10 & 8 & 6 & 100\\
 Deck 6   & 8 & 10 & \commonsuit 12 & \goalsuit 10 & 6 & 100\\
 Deck 7   & 10 & 8 & \commonsuit 12 & \goalsuit 10 & 6 & 100\\
 Deck 8   & 10 & 10 & \commonsuit 12 & \goalsuit 8 & 5 & 120\\
 Deck 9   & 8 & 10 & \goalsuit 10 & \commonsuit 12 & 6 & 100\\
 Deck 10  & 10 & 8 & \goalsuit 10 & \commonsuit 12 & 6 & 100\\
 Deck 11  & 10 & 10 & \goalsuit 8 & \commonsuit 12 & 5 & 120\\

\end{tabular}

\caption{Composition of each of the 12 possible compositions of a Figgie card deck, as well as the number of suits required to reach a majority and the associated payoff for doing so.  Note that for each deck, the yellow highlighted suit corresponds to the common suit while the green highlighted suit corresponds to the goal suit.  }
\label{table:decks}
\end{table}

\begin{algorithm}[H]
\DontPrintSemicolon
    Get two expected values $p_b$ and $p_s$, for buying and selling asset $j$ respectively.\;
    Randomly choose to buy or sell, with probability 0.5\;
    \eIf{buying}{
        Get a price $p$ according to Uniform$(0,p_b)$\;
        Let $s$ be the lowest sell order price in the market for asset $j$\;
        Send a limit buy order for asset $j$ with price $\min(p,s)$
    }{
        Get a price $p$ according to Uniform$(p_s,2p_s)$\;
        Let $b$ be the highest buy order price in the market for asset $j$\;
        Send a limit sell order for asset $j$ with price $\max(p,b)$
    }
 \caption{Order-sending based on expected value for asset $j$}
 \label{algo:order}
\end{algorithm}
\noindent Note that, for all trading strategies designed but the fundamentalist, $p_b=p_s$.

\paragraph{Noise Trader}

The noise trader simply computes its expected value for an asset according to:
\[p^*_j = b_j e^Z, \quad Z\sim N(0, \sigma^2)\]
Where $p^*_j$ is its expected value for asset $j$, $b_j$ is the price of the highest-price buy order in asset $j$'s order book, and $\sigma^2$ is the noise trader's variance factor, a parameter of trader that is set to $1$ by default.

\paragraph{Fundamentalist}

This agent uses different expected values dependent on whether it is to buy or sell a card, and determines these values by calculating a posterior distribution given the total number of distinct cards it has seen throughout the game as well as the number of cards of each suit it currently possesses. Algorithm \ref{algo:counting} describes how a fundamentalist can keep track of the number of distinct cards seen of a particular suit.

\begin{algorithm}[htp]
\DontPrintSemicolon
 Set $n$ to how many units of asset $j$ you are initially dealt\;
 Initialize a 4 element list $L$, such that for $x$ = self.num we set $L[x] = n$, and all other elements of $L$ are 0\; 
 Let $T$ be the list of trades for asset $j$, ordered by time\;
 \For{$i\gets 0$ \KwTo $|T|-1$}{
 \tcc{Since $T$ only grows, in practice we can store the data persistently and just run the for-loop for the new trades in $T$.}
  Let $b$ be the buyer in trade $T[i]$, and $s$ be the seller\;
  Let $v$ be the volume\;
  \eIf {$L[s$.num $] < v$}{
    Add $v$ to $L[b.$num$]$\;
    Set $L[s$.num$]$ to 0\;
  }{
    Add $v$ to $L[b.$num$]$\;
    Subtract $v$ from $L[s$.num$]$\;}
 }
 Return $L$
 \caption{Card-counting method for asset $j$, returning a list of known asset cards each agent holds}
 \label{algo:counting}
\end{algorithm}

\begin{wrapfigure}{R}{0.5\textwidth}
\centering
\vspace{-1\intextsep}
\setlength{\figW}{0.45\textwidth}
\input{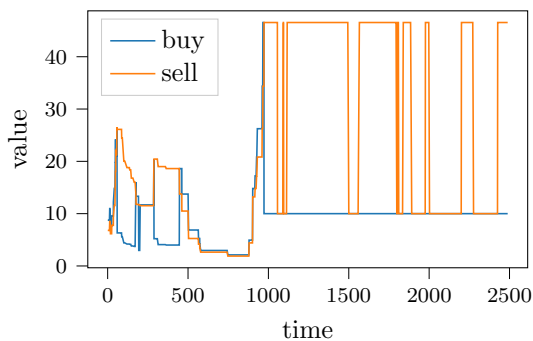}
\setlength{\abovecaptionskip}{0pt}
\caption{Note how the expected values change over time, particularly during the second half of the simulation. For this reason, clearing trades which no longer accurately reflect the expected value is essential to prevent the agent from performing poorly}
\label{figure:ev change}
\label{fig:fun_explanation}
\end{wrapfigure}

Using these tallies for each agent, the fundamentalist can then add the values of a given suit for each agent together to get the total number of distinct cards which it has seen throughout the whole game for each distinct suit.  Then, given this collection of cards, the agent calculates the total number of possible combinations by which it could be formed for each of the 12 decks.  Finally, a distribution of likelihoods for the possible deck is computed by dividing each of these numbers by their total sum, and the multinomial distribution $m$ is represented as follows:
\[ [m_0, m_1, m_2, m_3, m_4, m_5, m_6, m_7, m_8, m_9, m_{10}, m_{11}]\]
where $m_i$ is the probability that the deck being used in the game is deck $i$ as described in Table 1.

Unlike other traders, the fundamentalist agent has two different expected values for an asset - one for buying and one for selling.  Intuitively this makes sense, as the number of cards a player possesses ultimately affects the value of each card individually.  As an example, if a player possesses 6 spades, the most they can expect to gain from the purchase of an additional spade is 10 as they already are guaranteed a majority, while the sale of a spade could cost them up to 60 (loss of half the pot) plus an additional 10 in the event that spades is the goal suit.  Likewise, if a player possessed only 4 spades, they would be willing to buy at a higher price than if they possessed 6 (i.e. already were guaranteed a majority).  Thus, the expected value of a card must depend both on whether the agent is buying or selling that particular suit, as well as the number of cards of said suit it possesses.

We next introduce some notation.  Allowing spades, clubs, hearts, and diamonds to be labelled as assets 0, 1, 2 and 3 respectively, we have that given any suit $j$ that decks $3j$, $3j+1$, and $3j+2$ are the only ones in which the suit is the goal suit.  We define expected value functions $e_b(j, j_n, m)$ and $e_s(j, j_n, m)$ representing the expected values for buying and selling a suit $j$ respectively, given that the agent currently possesses $j_n$ cards of that suit and the probability distribution for the deck in play is $m$, as follows:
\\ 

\[ e_b(j, j_n, m) = \sum_{i=0}^{11} m_iv(i, j, j_n)\]
\[ e_s(j, j_n, m)= e_b(j, j_n - 1, m)\]
where $v(i, j, j_n)$ is a measure of the value of purchasing a card of suit $j$ assuming the deck in play is play $i$ and you have $j_n$ cards of suit $j$ in your hand.  Furthermore, $e_s$ is calculated as above due to the fact that intuitively, the expected value one gains by purchasing a card of a suit one possesses $j_n$ cards of is equivalent to the expected value one loses by selling a card of said suit when one possesses $j_n+1$ of these cards.  Calculating $v(i, j, j_n)$ is done as follows:

\[v(i, j, j_n) =
    \begin{cases}
        0 & \text{if } i \neq \ 3j, 3j+1, 3j+2  \\
        10 + v_m(i, j_n) & \text{if } i = \ 3j, 3j+1, 3j+2 \\
    \end{cases}
\]
where $v_m$ is the portion of a card's value arising from the final payout by way of possessing majorities (or tying for majority), which is given by: 

\[v_m(i, j_n) =
    \begin{cases}
        \frac{p_i(1 - r)}{1 - r^{x_i}}r^{j_n} & \text{if } j_n < x_i\\
        0 & \text{if } j_n \geq x_i\\
    \end{cases}
\]
where $x_i$ represents the number of goal suit cards needed to guarantee a majority for deck $i$, and $p_i$ represents the payout for doing so.  The constant $r$ is a parameter that is agent specific, and is such that if the price of purchasing a certain card of a suit $j$ when an agent possesses 0 cards of this suit is $a_i$, then the value towards obtaining a majority of purchasing a card of this suit given the agent currently possesses $n$ such cards is $a_ir^n$ for some $r >1$ when $n < x_i$, and is 0 otherwise. The reason this valuation was decided upon was due to the fact that as an agent gets closer to reaching a majority, the more likely they will be able to successfully be able to obtain it, and hence the more they should be willing to spend in doing so.

The value of $a_i$ is computed for each deck $i$ such that:
\[ p_i = a_{i} + a_{i}r + \ldots + a_ir^{x_i-1}\]
This way, the portion of the expected value associated with buying each card, assuming an agent started with 0, would sum to the total value gained from the payout for a majority.  Solving for $a_i$, we find that:
\[ a_i = \frac{p_i(1 - r)}{1 - r^{x_i}}\]
as desired.

Using the above equations, the algorithm which the fundamentalist agent uses is shown in Algorithm \ref{algo:fundamentalist} below. 

\begin{algorithm}[htp]
\DontPrintSemicolon
 Let $m$ be the multinomial distribution obtained by card counting and the methods detailed above \;
 Initialize a two element list $L = [ e_b(j, j_n, m), e_s(j, j_n, m)]$\;
 Let $O_b$ and $O_s$ be the agent's own order books for buy and sell orders respectively for asset $j$\;
 \For{$i\gets 0$ \KwTo $|O_b|-1$}{
  Let $o$ be the order $O_b[i]$\;
  \If {$o$.price$>L[0]$}{
    Let $o$.deleted be $True$\;
  }
 }
 
 \For{$i\gets 0$ \KwTo $|O_s|-1$}{
  Let $o$ be the order $O_b[i]$\;
  \If {$o$.price$<L[1]$}{
    Let $o$.deleted be $True$\;
  }
  }
 Return $L$
 \caption{Fundamentalist trading algorithm for asset $j$}
 \label{algo:fundamentalist}
\end{algorithm}

Note that upon each iteration, all orders currently in the market that no longer agree with the agent's expected values are cleared by way of lazy deletion.  Each agent has an associated order book for buy orders as well as an associated order book for sell orders for each asset which contains the orders which it has placed.  Lazy deletion allows for efficient deletion of these orders (as compared to removing them from the min and max-heaps directly).  Furthermore, doing so is necessary, as the agent's expected values can fluctuate significantly throughout the game, as seen in Figure \ref{figure:ev change} below, due to the fact that the price is a reflection not only of the belief state of the deck, but of the cards the player currently possesses.

\paragraph{Bottom-Feeder} The bottom-feeder is a social trader. It calculates its expected value for an asset by estimating that of other agents (its ``prey'') from their order histories, and averaging them. When looking at an agent, it looks at its past $k$ buy order prices and past $k$ sell order prices and computes their mean. In our model, we used $k=4$. Effectively, the bottom-feeder calculates the expected value for asset $j$ according to:
\[p_j^* = \frac{1}{|A|}\sum_{i\in A} \frac{1}{2}\left(\frac{1}{k} \sum_{p_b \in B_{ji}(k)}p_b+\frac{1}{k} \sum_{p_s \in S_{ji}(k)}p_s\right),\]
where $p_j^*$ is the bottom-feeders expected value for asset $j$, $A$ is the set of prey, $k$ is how many buy and sell orders the bottom-feeder will examine for each agent, and $B_{ji}(k)$ and $S_{ji}(k)$ are sets of the $k$ most recent buy and sell orders respectively of agent $i$ for asset $j$. This strategy produces good estimates of an agent's expected value for an asset under the assumption that their buy and sell orders are, on average, the same distance from the agent's expected value. Note that, in practice, if an agent the bottom-feeder is looking at has not sent $k$ buy orders and $k$ sell orders for an asset, it is not included in $A$.

\paragraph{Chartist}
The chartist trader only looks at the price series data to estimate an expected value for an asset. It does this by estimating the average rate of return of an asset and assuming that it will continue. For example, if the average rate of return is high, then the chartist will be willing to buy the asset for a high price as the price in the future should be even higher.

There are several ways to estimate an expected return. Chiarella et. al. give the following formula \cite{heterogenous}:

\[\bar{r}_t^i = \frac{1}{\tau^i}\sum_{j=1}^{\tau^i}r_{t-j}=\frac{1}{\tau^i}\sum_{j=1}^{\tau^i}\ln\frac{p_{t-j}}{p_{t-j-1}},\]
where $\bar{r}_t^i$ is the average spot return, $\tau^i$ the chartist's time horizon (i.e. how far back in the data the chartist will look), $r_{t-j}$ is the spot return at time $t-j$, and $p_{t-j}$ is the price at time $t-j$. The estimated future value using this method is then $p_t\exp(\bar{r}_t^i \tau^i)$ \cite{heterogenous}. Note that the logarithm telescopes such that:
\begin{align}\bar{r}_t^i &= \frac{1}{\tau^i}\sum_{j=1}^{\tau^i}\ln\frac{p_{t-j}}{p_{t-j-1}} = \frac{1}{\tau^i}\sum_{j=1}^{\tau^i}\ln(p_{t-j})-\ln(p_{t-j-1}) \nonumber\\
\bar{r}_t^i &= \frac{1}{\tau^i} \ln \frac{p_{t-1}}{p_{t-\tau^i-1}},\end{align}\label{chartist1}
making the formula effectively discount all of the data between the current price and the price at the beginning of the time horizon. 

Another option is to compute a linear regression on a part of the time series, according to the model:
\begin{equation}\label{chartist2}
\ln(p_t) = y_t = \beta_0 + \beta_1 t + \epsilon,
\end{equation}
on a set of times and their associated prices within the time horizon of the chartist. The estimated future value is then $\hat{p_t} = \exp(\hat\beta_0 + \hat\beta_1t)$.

\section{Results}
In this section, we are going to demonstrate the performance of our agents and evaluate our model using statistics. Note all bar graphs in this section use the same bar color and labeling for each kind of agent: 
\begin{itemize}
    \item Fundamentalist: blue bars with label (f, f0, f1, ...)
    \item Bottom-feeder: orange bars with label (b, b0, b1, ...)
    \item Noise trader: green bars with label (n, n0, n1, ...)
\end{itemize}

\subsection{Strategy performance}
We run the simulation 100 times for each possible combination of the three types of agents. The hight of a bar is average final gain of an agent with the line in the middle of each bar representing mean final cash. Bars are $\pm$2SE, with the grey ones being the standard error of the mean cash and the black ones being the error of the final gain. The grey dotted line represents beginning cash level ($\$350$ per agent) and the black dotted line ($ \$400$) is the average bonus of the game (since each player put $\$50$ into the pot at the beginning of each round) We observe that noise trader's overall performance is worse than the other two in any simulation setup.

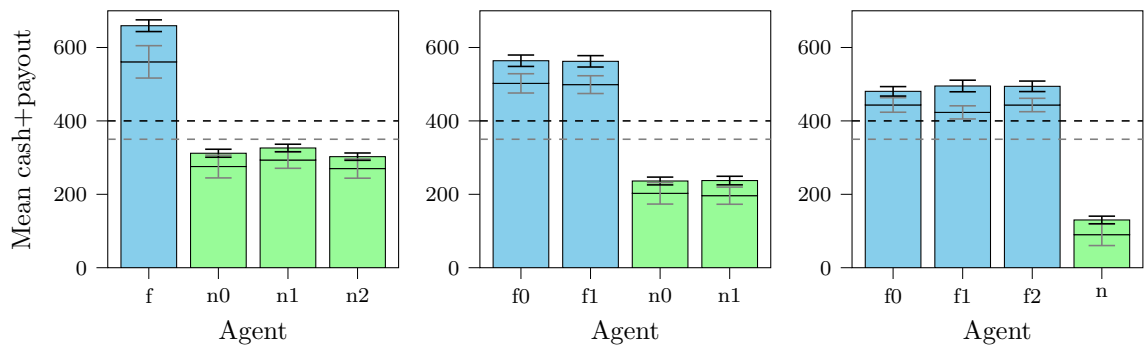
\begin{figure}[H]
\setlength{\figW}{0.33\textwidth}
\centering
\input{Graphs/1f3n}
\input{Graphs/2f2n}
\input{Graphs/3f1n}
\setlength{\abovecaptionskip}{0pt}
\caption{100 simulations each for 3 setups of fundamentalists against noise traders.}
\label{fig:funvsnoisy}
\end{figure}

\paragraph{Fundamentalist} When there are only fundamentalists and noise traders involved in the simulation (see Figure \ref{fig:funvsnoisy}, we observe the fundamentalists earn more, on average, compared to noise traders. We observe the fundamentalist also have a much higher standard deviation than the noise trader, mainly due to unsuccessful guesses of the actual goal suit in some rounds. However, the overall performance of fundamentalists is much better than noise traders, with the general trend that the more fundamentalists exist in a game the less the average gain is per fundamentalist due to increased market competition between themselves.

\paragraph{Bottom-Feeder} In our model, the bottom-feeder chooses one or more  agents as its prey. We discover that when all bottom-feeders choose other bottom-feeders as its prey, it will not make any trade. Bottom-feeders only make trade after observing its prey's trade, when all bottom-feeders are looking after each other, no trade will be made. Thus, we will only consider the cases when at least one bottom-feeder chooses another type of agent as prey.

When bottom-feeder is added into the fundamentalist - noise trader system, results are pretty different depending on its prey. When using the fundamentalist as its prey (see Figure \ref{fig:one_bf_fun} and Figure \ref{fig:two_bf_fun}, the bottom-feeder outperform the noise trader, but gains less on average than the fundamentalist. Since the bottom-feeder is only following the fundamentalist, it is always one step slower. Thus, the resulting average cash and payout for the fundamentalist in this case is actually higher than a fundamentalist in two-fundamentalist-two-noise-trader case shown in Figure \ref{fig:funvsnoisy}.

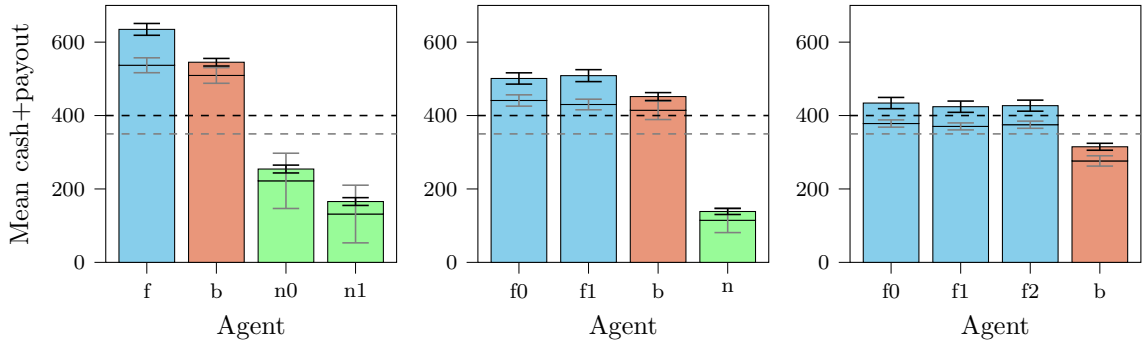
\begin{figure}[H]
\setlength{\figW}{0.33\textwidth}
\centering
\input{Graphs/bvsf.1}
\input{Graphs/bvsf.2}
\input{Graphs/bvsf.3}
\setlength{\abovecaptionskip}{0pt}
\caption{One bottom-feeder with all fundamentalists in the game as preys. (We discovered there is no statistical significance between using one fundamentalist as prey or all fundamentalists in the game as prey, so only demonstrating one case here.)} \label{fig:one_bf_fun}
\end{figure}

\begin{figure}[H]
\centering
\setlength{\figW}{0.33\textwidth}
\input{Graphs/2bf_fun.1}
\input{Graphs/2bf_fun.2}
\setlength{\abovecaptionskip}{0pt}
\caption{Two bottom-feeders with all  fundamentalists in the game as preys.} \label{fig:one_bf_fun}
\end{figure}
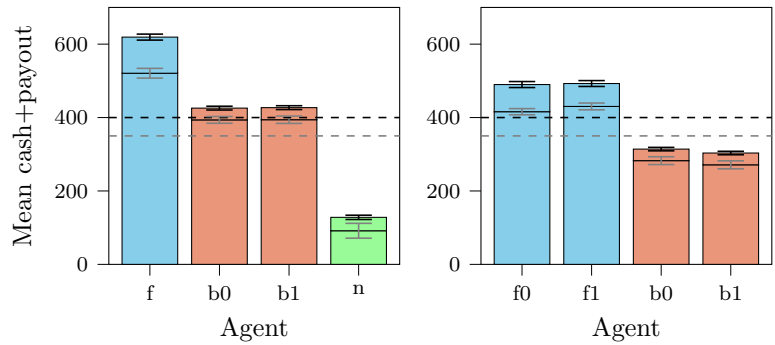

When the bottom-feeder uses the noise trader as its prey, it gains similar average cash but more bonus than the noise trader. 
Since the bottom-feeder tries to follow its prey's expectations, we expect it perform a little worse than its prey. However, when put into three-bottom-feeder-one-noise-trader system (see Figure\ref{fig:three_bf_n}) with the noise trader as its prey, the bottom-feeders actually outperforms the noise trader.

\begin{figure}[H]
\setlength{\figW}{0.5\textwidth}
\includegraphics[width=\textwidth]{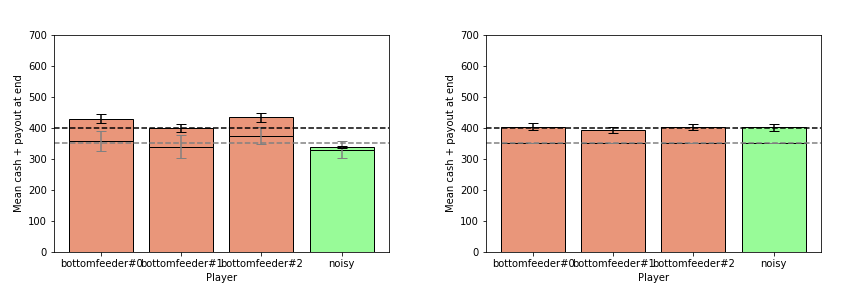}
\setlength{\abovecaptionskip}{0pt}
\caption{Three bottom-feeders with the left using noise trader as prey and right using itself as prey (Ex: b0 is using b0 as prey and b1 is using b1 as prey).} \label{fig:two_bf_n}
\end{figure}

\begin{figure}[H]
\includegraphics[width=\textwidth]{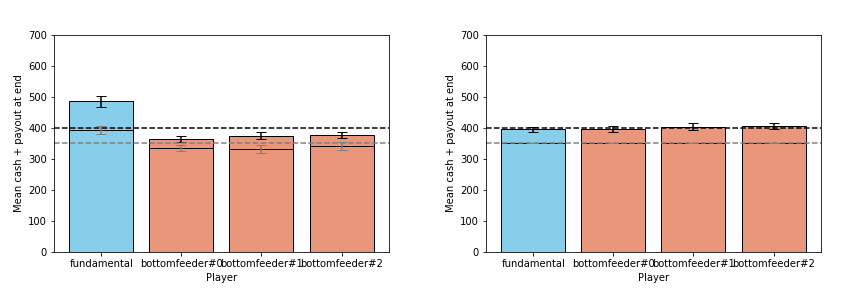}\setlength{\abovecaptionskip}{0pt}
\caption{Three bottom-feeders with the left using fundamentalist as prey and right using itself as prey.} \label{fig:three_bf_fun}
\end{figure}

\paragraph{Chartist}
Neither of the algorithms implemented for the chartist's estimation of future expected value functioned within the simulation framework. Both produced estimates extreme enough to be too large to store in double-precision floating point numbers. The reasons for this will be discussed later.

\begin{wraptable}{R}{0.5\textwidth}
    \centering
    \caption{Standard deviations of Noise, Fundamentalist, and Bottom-feeder strategy final wealth in $n=100$ games each where all 4 players are of one strategy. All one-tailed tests between standard deviations of each category have $p<10^{-5}$, except that the standard deviation of the fundamentalists' payouts are greater than that of the noise traders with $p=0.013$\iffalse Variance of noise traders is greater than that of fundamentalists with $p=1.20\times 10^{-102}$, and greater than that of bottom-feeders with $p=2.08\times 10^{-138}$. Variance of fundamentalists is greater than that of bottom-feeders with $p=9.95 \times 10^{-7}$\fi}
    \begin{tabular}{cccc}
         & \multicolumn{3}{c}{SD}\\
         \cline{2-4}
        \textbf{Strategy} &  Final payout & Cash & Bonus\\ \hline
        Noise & 201.04 & 188.52 & 61.36\\
        Fundamentalist & 61.75 & 33.11 & 68.62\\
        Bottom-feeder & 48.63 & 0 & 48.63\\\hline
    \end{tabular}
    \label{tab:puresds}
    \vspace{-2\intextsep}
\end{wraptable}

\paragraph{Homogeneous set-ups}
When the simulation is run with all agents being of one type with the same parameters, the final payoff, cash, and bonus for each agent are \$400, \$350, and \$50 respectively. This is obvious, as agents with identical strategies cannot, on average, outperform each other. But what is not obvious is how much variability there is in the overall payout, cash, and bonus of each agent in these homogeneous simulations. Table \ref{tab:puresds} describes the standard deviations of the three. The standard deviation on the bottom-feeders' cash is 0 because, in a game with four bottom-feeders, no trades occur as their trading behavior requires observing trades, as mentioned in the bottom-feeder paragraph in this section.

\subsection{Expectations}
\begin{figure}[H]
\includegraphics[width=\textwidth]{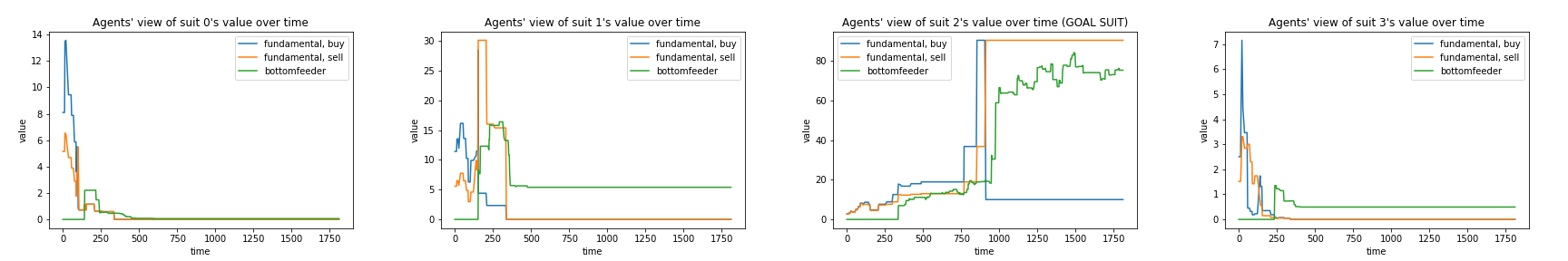}\setlength{\abovecaptionskip}{0pt}
\caption{Agents' Expectations of Each Suit Over Time} \label{fig:1bf_fun_expectations}
\end{figure}

\begin{figure}[H]
\includegraphics[width=\textwidth]{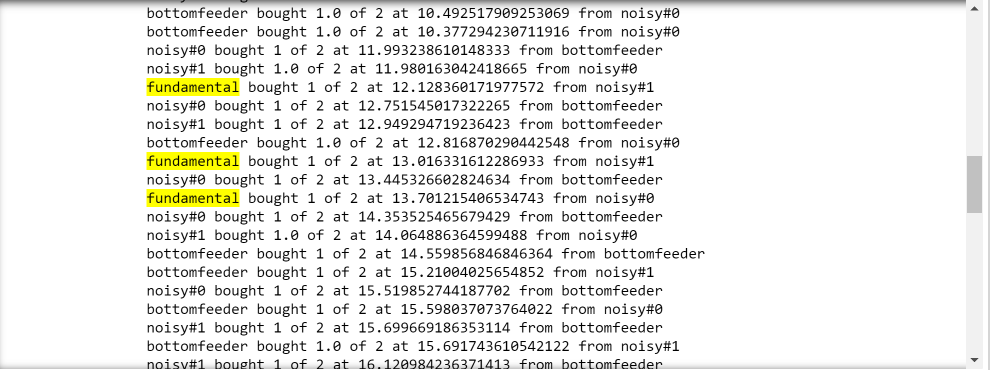}\setlength{\abovecaptionskip}{0pt}
\caption{Agents' Expectations of Each Suit Over Time} \label{fig:fun_purhchase}
\end{figure}

An agent's expectation of an asset during one round of Figgie is determined by its algorithm as introduced in Section 2.3. Since the noise trader's expectation is random, we will focus only on the fundamentalist and bottom-feeder in this section. Theoretically, the fundamentalist should have a guess on the goal suit after a few trades have been made, therefore its expectation for all other three suits will drop to zero except for the potential goal suit, as shown in Figure \ref{fig:1bf_fun_expectations}. In this round of simulation, the fundamentalist correctly guessed the goal suit. Its expected buy price rose to over 80 because the fifth goal suit card will not only bring the \$10 bonus, but also the \$100 extra bonus at the end. After that peak , the expected buy price for the fundamentalist dropped to \$10, which is exactly what the fundamentalist can get out of this additional goal suit card. Notice the expected sell price for the goal suit card skyrocketed to over \$80 for the fundamentalist at the same time as the peak in buying price, because selling that additional goal suit card (the sixth one) means the fundamentalist has the potential to lose half the \$100 bonus. The selling price remaining high represents the fundamentalist ending up with exactly five goal suit cards until the end of this game. Otherwise the price would drop since the seventh card is not worth as much to the fundamentalist as the sixth card because of the majority rule in Figgie. In this round of simulation, the bottom-feeder's prey is the fundamentalist so that we can better observe the relationship between the two agents' expectations. We can observe that the bottom-feeder 's expected price for suit 2 rise after the sudden peak of the fundamentalist's expectations. By investigating all the trades made in this simulation, we can see that the fundamentalist bought suit 2 three times at a relatively high price, above \$10. Notice since \$10 is the bonus for owning one goal suit card, buying a card at prices above \$10 means this agent is pretty confident that this suit is the goal suit. After observing the fundamentalist's actions, the bottom-feeder's expected price also rises to about \$10 (see Figure \ref{fig:fun_purhchase}). 

\subsection{Latencies}

As discribed in Section 2.1, to test the effect of latencies on agents' performance, we run a homogeneous simulation filled with fundamentalists and give one of them (\textit{fast}) a latency of 0 and three (\textit{slow, slow1} and \textit{slow2}) a latency of 100. The results are displayed in Figure \ref{fig:speeds_rewards}, which displays the rewards, and Table \ref{tab:speeds_bootstrap}, which displays confidence intervals. Note that the confidence intervals are bootstrapped intervals of the difference between two means. The bootstrap is used because:
\begin{itemize}
    \item Figgie is a zero-sum game, so agents' wealth, cash, and payout are not independent. This precludes tests that identify whether two sets of independent variables have different means.
    \item We find that the distributions of wealth, cash, payout, and of their differences between agents, are bimodal. This again makes it difficult to conduct standard statistical tests.
\end{itemize}

\begin{figure}[H]
\setlength{\figW}{0.5\textwidth}
\centering
Cash+payout of agents with different latencies\par\medskip
\input{Graphs/speeds}
\setlength{\abovecaptionskip}{0pt}
\caption{Cash and payout of one fast ($\lambda=0$) and three slow ($\lambda=100$) agents. All are fundamentalists. The fast agent tends to finish with less cash than the others, but appears to receive slightly more of the payout on average. We speculate that is because the fast agent made several unsuccessful trades at the beginning of the simulation.} \label{fig:speeds_rewards}
\end{figure}
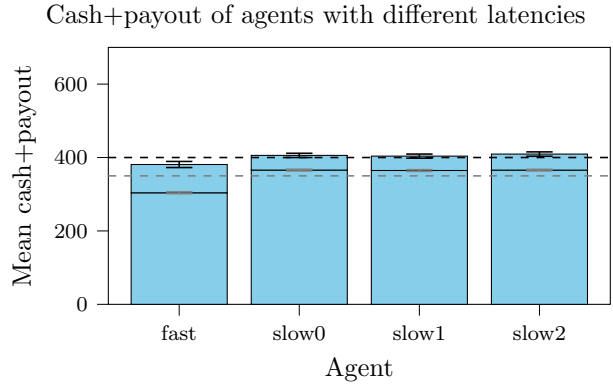
\begin{table}[H]
\centering
\begin{tabular}{ccccccc}
     & \multicolumn{2}{c}{$\Delta$Wealth} & \multicolumn{2}{c}{$\Delta$Cash} & \multicolumn{2}{c}{$\Delta$Payout}\\
     \cline{2-3}\cline{4-5}\cline{6-7}
    \textbf{Competitor} &  2.5\% & 97.5\% & 2.5\% & 97.5\% & 2.5\% & 97.5\% \\ \hline
    slow0 & 15.67 & 38.44 & 58.14 & 65.84 & -46.75 & -22.95\\
    slow1 & 11.63 & 33.99 & 57.22 & 64.83 & -49.77 & -26.35\\
    slow2 & 17.98 & 40.74 & 58.30 & 65.82 & -44.35 & -20.32\\\hline
\end{tabular}
\caption{Bootstrapped 95\% confidence intervals on the mean difference between \textit{fast} and its competitors' wealth, cash, and payout and. Positive values indicate that the competitor had more than \textit{fast}. In all but payout, the slower agents, on average, receive more than \textit{fast}. In payout, \textit{fast} tends to receive more.}
\label{tab:speeds_bootstrap}
\end{table}

\subsection{Simulation statistics}
\begin{figure}[H]
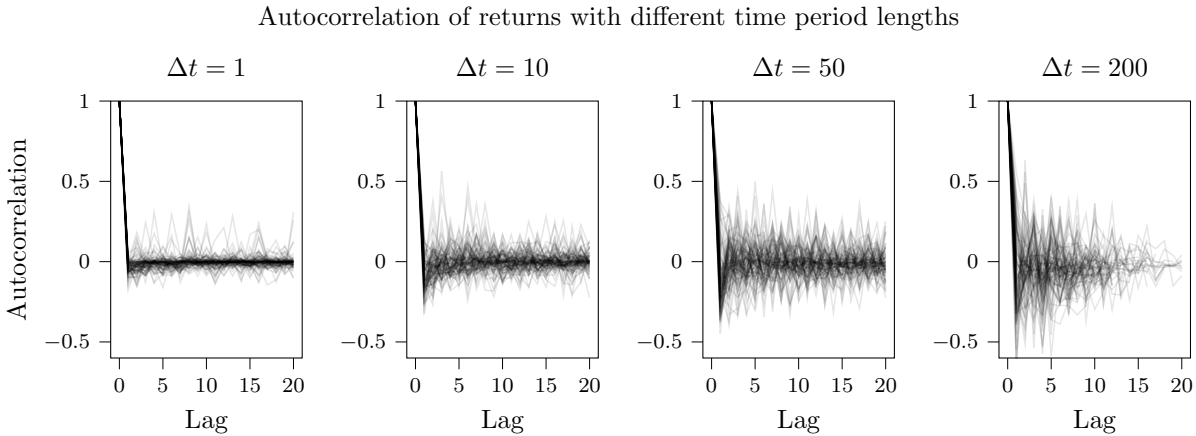

\setlength{\figW}{0.25\textwidth}
\centering
Autocorrelation of returns with different time period lengths\par\medskip
\input{Graphs/acf/acf.1}
\input{Graphs/acf/acf.10}
\input{Graphs/acf/acf.50}
\input{Graphs/acf/acf.200}
\setlength{\abovecaptionskip}{0pt}
\caption{Autocorrelation of the returns from 100 simulations, each of which is represented by a grey line. Taken from simulations with one fundamentalist and three noise traders. Autocorrelations are centered around 0, getting more extreme with longer time period lengths.} \label{fig:acfperiod}
\end{figure}

\begin{figure}[H]
\centering
\setlength{\figW}{0.5\textwidth}
\input{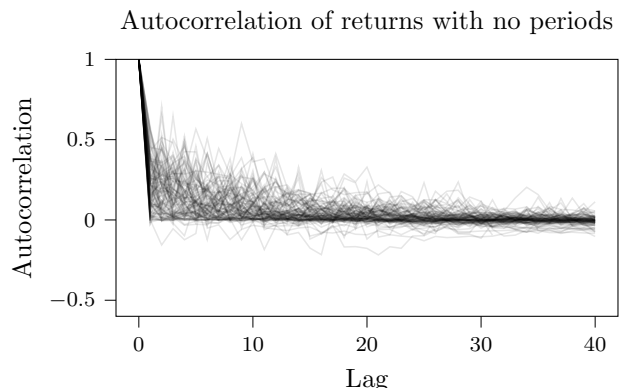}
\setlength{\abovecaptionskip}{0pt}
\caption{Autocorrelation of the returns from 100 simulations, each of which is represented by a grey line. Taken from simulations with one fundamentalist and three noise traders}
\label{fig:acfnoperiod}
\end{figure}

\paragraph{Auto-correlation}
The auto-correlation function is often used on the series of returns of a particular asset, where a return is the ratio of the closing price of one trading day to that on the previous trading day. Since our simulation has no concept of trading periods, we can arbitrarily set periods and observe the behavior of the autocorrelation. The autocorrelation of the return series for some time periods is shown in Figure \ref{fig:acfperiod}. Figure \ref{fig:acfnoperiod} shows the autocorrelation of the entire return series, without setting any time periods.

\section{Discussion}

\paragraph{Success in simulating the fundamentalist} Fundamentalist trades according to its own expected value of the asset, and we have successfully implemented the card-counting method for each asset for the fundamentalist such that its overall performance is satisfying. When put into a market with other kinds of traders, as shown in Figures \ref{fig:funvsnoisy}, \ref{fig:one_bf_fun}, and \ref{fig:two_bf_fun} in the results section, it can always land in at least the same average cash level as the beginning cash (\$350), and its ending bonus is also higher than both the noise trader and the bottom-feeder. The average cash for the fundamentalist of all cases shown in the result section is \$407.5 and average bonus is \$75.43.

\paragraph{Possible Improvements for the bottom-feeder} Currently our bottom-feeder's prey is manually chosen before the start of each simulation, thus its according performance will totally based on its prey. In future development of this model, we plan to add an algorithm that enables the bottom-feeders to pick its own prey based on each agent's performance it observes in one game. This is challenging since the total time of the game is relatively short, thus not much data is collected. But if the model is expanded and the time for each round is extended, this will be a meaningful improvement.

\paragraph{Failure of chartist trading strategies}
There are two key differences between our market model as applied to Figgie and previous work that, combined, may be the reason for the failure of chartist strategies.
\begin{enumerate}
    \item \textbf{Small market:} Figgie is, by nature, a very small game. It contains just four traders and four assets, and trades happen on the order of dozens or hundreds, not thousands or millions. This means that, if the chartist is using Equation \eqref{chartist1}, it is likely that the current price is the price of the chartist's most recent buy order. If a chartist slightly over-estimates the asset's value, and then uses that over-estimate again to define the most recent price and calculate a new estimate, the new estimate will be even higher. This can evidently go on until the estimated value becomes absurd. The linear regression model in Equation \eqref{chartist2} is slightly more robust, as it takes into account all of the recent prices, but overestimates create orders which create outliers in the price series, which then increase the estimates, etc.
    \item \textbf{Risk ignored:} As agents are trying to maximize their average returns, risk is ignored by the agents in our simulation. This removes one method through which chartists' price estimates are moderated downwards, exacerbating the problem of an overestimation feedback loop.
\end{enumerate}
It is unclear how to make a chartist trading strategy that ignores risk and functions in a small market without creating extreme estimates. One option is to have it exclude its own orders when calculating the current price. However, this may not work, as noise traders' orders will follow the market price, which, for a small moment, is set by the chartist. Note also that, in line with previous work, the chartists did increase volatility in the market. But the volatility induced by a single chartist in the small market was too great for the limitations of double-precision floating-point numbers.

\paragraph{Autocorrelations}
When arbitrarily setting time periods, Figure \ref{fig:acfperiod} shows that the autocorrelation of the returns is centered around 0. This indicates that a chartist strategy, even if a robust one were created, would not be able to generate returns if splitting the data into periods. Figure \ref{fig:acfperiod} also shows that the autocorrelation gets more extreme as the lengths of the time periods ($\Delta t$) increase. This is likely a result of an increase in $\Delta t$ reducing the number of time periods when holding the duration of the simulation constant. With just a few points in the series of returns, it is more likely for there to be a spurrious correlations between returns at various lags.

When not splitting time periods, as in Figure \ref{fig:acfnoperiod}, the autocorrelation of the returns is generally above 0 and decreases towards 0 as the lag increases. This indicates that there is room for a chartist strategy to exploit this correlation and generate returns, as seeing a positive/negative return would tell it that the next return is likely to be positive/negative, and it send make buy and sell orders accordingly. However, as noted above, the chartist strategies tested failed in Figgie due to feedback loops, so a more robust strategy would need to be developed to exploit this pattern.

\paragraph{Latency}
As shown in Table \ref{tab:speeds_bootstrap}, a lower latency relative to other fundamentalists decreases a fundamentalist's mean cash and wealth, but might increase its mean payout. The increase in mean payout is to be expected, as a trader with lower latency should be better able to acquire cards of the goal suit, and always trades with slightly more recent information about which suit might be the goal suit. The result that mean cash and wealth decrease is surprising, as the low-latency trader should be able to make the best possible trades as it always sees the current market prices. One possible explanation for this is that a lower latency means a higher trading frequency, because in the simulation agents will only start considering making a trade again after their order is added to the order book. This might make the trader make more unfavourable trades in the beginning of the game, when its estimated asset values fluctuate rapidly (see, for example, Figure \ref{fig:fun_explanation}).

Table \ref{tab:speeds_bootstrap} also demonstrates the robustness of our simulation and of the effectiveness of the bootstrap method. The confidence intervals on the differences in wealth, cash, and payout between \textit{fast} and \textit{slow0, slow1}, and \textit{slow2} are very close for all of \textit{fast}'s competitors. This shows that different instantiations of identical agents in our simulation do indeed perform the same, on average. The similarity of the confidence intervals also shows that the bootstrap method, even with just 100 simulations, can be used to statistically compare strategies in Figgie.

\section{Conclusion}

In this paper, we have developed a general agent-based discrete-event simulation for modelling trading dynamics in a market.  Our simulation is flexible enough to extend far beyond the Figgie setting, allowing for various market conditions including number of traders, assets, and latencies to be set as parameters and the easy user testing of new trading algorithms.  One innovation of our simulation is the efficient implementation of latencies between traders and the market. While this method is unable to simulate latencies between traders, it is enough to capture a market where traders do not directly communicate with each other.

We have designed and implemented a few different trading strategies for Figgie, with our fundamentalist strategy taking advantage of its unique payout structure and our ``bottom-feeder'' strategy taking advantage of its nature as a market where everything is visible. We have also implemented a chartist trading strategy from the literature and designed our own. The failure of the chartist strategies demonstrates that, if chartists ignore risk and operate in a sufficiently small market, they may enter a feedback loop where high estimates cause high prices, which cause even higher estimates, etc. We recommend further research into the design of robust chartist trading strategies which do not fall into feedback loops, as well as the exact conditions required for feedback loops to occur.

We have demonstrated the utility of the bootstrap procedure in statistically comparing the outcomes of strategies in zero-sum games, in which traditional statistical tests fail due to the nonstandard distributions that arise and the dependence of outcomes between agents.

The discrete-event nature of our simulation allows us to study how the autocorrelation of an asset's returns differs when setting different trading period lengths and when not using periods at all. This ability demonstrates an advantage of discrete-event simulation over market simulations where time passes in fixed steps.

\bibliographystyle{plain}
\bibliography{references}

\end{document}

%% file: Graphs/1f3n.tex
\begin{tikzpicture}

\definecolor{color0}{rgb}{0.529411764705882,0.807843137254902,0.92156862745098}
\definecolor{color1}{rgb}{0.596078431372549,0.984313725490196,0.596078431372549}

\begin{axis}[
height=\figH,
tick align=outside,
tick pos=left,
width=\figW,
x grid style={white!69.0196078431373!black},
xlabel={Agent},
xmin=-0.59, xmax=3.59,
xtick style={color=black},
xtick={0,1,2,3},
xticklabels={f,n0,n1,n2},
y grid style={white!69.0196078431373!black},
ylabel={Mean cash+payout},
ymin=0, ymax=700,
ytick style={color=black}
]
\draw[draw=black,fill=color0] (axis cs:-0.4,0) rectangle (axis cs:0.4,560.827464238735);
\draw[draw=black,fill=color1] (axis cs:0.6,0) rectangle (axis cs:1.4,275.678587219141);
\draw[draw=black,fill=color1] (axis cs:1.6,0) rectangle (axis cs:2.4,293.410060108814);
\draw[draw=black,fill=color1] (axis cs:2.6,0) rectangle (axis cs:3.4,270.08388843331);
\draw[draw=black,fill=color0] (axis cs:-0.4,560.827464238735) rectangle (axis cs:0.4,659.327464238735);
\draw[draw=black,fill=color1] (axis cs:0.6,275.678587219141) rectangle (axis cs:1.4,311.878587219141);
\draw[draw=black,fill=color1] (axis cs:1.6,293.410060108814) rectangle (axis cs:2.4,326.210060108814);
\draw[draw=black,fill=color1] (axis cs:2.6,270.08388843331) rectangle (axis cs:3.4,302.58388843331);
\path [draw=white!50.1960784313725!black, semithick]
(axis cs:0,516.70959360209)
--(axis cs:0,604.94533487538);

\path [draw=white!50.1960784313725!black, semithick]
(axis cs:1,244.670242463879)
--(axis cs:1,306.686931974402);

\path [draw=white!50.1960784313725!black, semithick]
(axis cs:2,270.887704658008)
--(axis cs:2,315.932415559621);

\path [draw=white!50.1960784313725!black, semithick]
(axis cs:3,243.922319166524)
--(axis cs:3,296.245457700096);

\path [draw=black, semithick]
(axis cs:0,643.379110091676)
--(axis cs:0,675.275818385793);

\path [draw=black, semithick]
(axis cs:1,300.979853347722)
--(axis cs:1,322.777321090559);

\path [draw=black, semithick]
(axis cs:2,315.972560414065)
--(axis cs:2,336.447559803564);

\path [draw=black, semithick]
(axis cs:3,292.582388545793)
--(axis cs:3,312.585388320827);

\addplot [semithick, white!50.1960784313725!black, mark=-, mark size=5, mark options={solid}, only marks]
table {%
0 516.70959360209
1 244.670242463879
2 270.887704658008
3 243.922319166524
};
\addplot [semithick, white!50.1960784313725!black, mark=-, mark size=5, mark options={solid}, only marks]
table {%
0 604.94533487538
1 306.686931974402
2 315.932415559621
3 296.245457700096
};
\addplot [semithick, black, mark=-, mark size=5, mark options={solid}, only marks]
table {%
0 643.379110091676
1 300.979853347722
2 315.972560414065
3 292.582388545793
};
\addplot [semithick, black, mark=-, mark size=5, mark options={solid}, only marks]
table {%
0 675.275818385793
1 322.777321090559
2 336.447559803564
3 312.585388320827
};
\addplot [semithick, white!50.1960784313725!black, dashed]
table {%
-0.59 350
3.59 350
};
\addplot [semithick, black, dashed]
table {%
-0.59 400
3.59 400
};
\end{axis}

\end{tikzpicture}

%% file: Graphs/2f2n.tex
\begin{tikzpicture}

\definecolor{color0}{rgb}{0.529411764705882,0.807843137254902,0.92156862745098}
\definecolor{color1}{rgb}{0.596078431372549,0.984313725490196,0.596078431372549}

\begin{axis}[
height=\figH,
tick align=outside,
tick pos=left,
width=\figW,
x grid style={white!69.0196078431373!black},
xlabel={Agent},
xmin=-0.59, xmax=3.59,
xtick style={color=black},
xtick={0,1,2,3},
xticklabels={f0,f1,n0,n1},
y grid style={white!69.0196078431373!black},
ymin=0, ymax=700,
ytick style={color=black}
]
\draw[draw=black,fill=color0] (axis cs:-0.4,0) rectangle (axis cs:0.4,502.243444038788);
\draw[draw=black,fill=color0] (axis cs:0.6,0) rectangle (axis cs:1.4,498.809267721344);
\draw[draw=black,fill=color1] (axis cs:1.6,0) rectangle (axis cs:2.4,202.669833794141);
\draw[draw=black,fill=color1] (axis cs:2.6,0) rectangle (axis cs:3.4,196.277454445726);
\draw[draw=black,fill=color0] (axis cs:-0.4,502.243444038788) rectangle (axis cs:0.4,563.943444038788);
\draw[draw=black,fill=color0] (axis cs:0.6,498.809267721344) rectangle (axis cs:1.4,562.309267721344);
\draw[draw=black,fill=color1] (axis cs:1.6,202.669833794141) rectangle (axis cs:2.4,236.269833794141);
\draw[draw=black,fill=color1] (axis cs:2.6,196.277454445726) rectangle (axis cs:3.4,237.477454445726);
\path [draw=white!50.1960784313725!black, semithick]
(axis cs:0,475.981643752754)
--(axis cs:0,528.505244324823);

\path [draw=white!50.1960784313725!black, semithick]
(axis cs:1,474.578261651598)
--(axis cs:1,523.04027379109);

\path [draw=white!50.1960784313725!black, semithick]
(axis cs:2,173.443784150446)
--(axis cs:2,231.895883437836);

\path [draw=white!50.1960784313725!black, semithick]
(axis cs:3,172.900458058793)
--(axis cs:3,219.654450832659);

\path [draw=black, semithick]
(axis cs:0,548.399815118874)
--(axis cs:0,579.487072958703);

\path [draw=black, semithick]
(axis cs:1,546.80862257348)
--(axis cs:1,577.809912869208);

\path [draw=black, semithick]
(axis cs:2,225.658443687279)
--(axis cs:2,246.881223901003);

\path [draw=black, semithick]
(axis cs:3,225.814590303501)
--(axis cs:3,249.140318587951);

\addplot [semithick, white!50.1960784313725!black, mark=-, mark size=5, mark options={solid}, only marks]
table {%
0 475.981643752754
1 474.578261651598
2 173.443784150446
3 172.900458058793
};
\addplot [semithick, white!50.1960784313725!black, mark=-, mark size=5, mark options={solid}, only marks]
table {%
0 528.505244324823
1 523.04027379109
2 231.895883437836
3 219.654450832659
};
\addplot [semithick, black, mark=-, mark size=5, mark options={solid}, only marks]
table {%
0 548.399815118874
1 546.80862257348
2 225.658443687279
3 225.814590303501
};
\addplot [semithick, black, mark=-, mark size=5, mark options={solid}, only marks]
table {%
0 579.487072958703
1 577.809912869208
2 246.881223901003
3 249.140318587951
};
\addplot [semithick, white!50.1960784313725!black, dashed]
table {%
-0.59 350
3.59 350
};
\addplot [semithick, black, dashed]
table {%
-0.59 400
3.59 400
};
\end{axis}

\end{tikzpicture}

%% file: Graphs/3f1n.tex
\begin{tikzpicture}

\definecolor{color0}{rgb}{0.529411764705882,0.807843137254902,0.92156862745098}
\definecolor{color1}{rgb}{0.596078431372549,0.984313725490196,0.596078431372549}

\begin{axis}[
height=\figH,
tick align=outside,
tick pos=left,
width=\figW,
x grid style={white!69.0196078431373!black},
xlabel={Agent},
xmin=-0.59, xmax=3.59,
xtick style={color=black},
xtick={0,1,2,3},
xticklabels={f0,f1,f2,n},
y grid style={white!69.0196078431373!black},
ymin=0, ymax=700,
ytick style={color=black}
]
\draw[draw=black,fill=color0] (axis cs:-0.4,0) rectangle (axis cs:0.4,443.457580821169);
\draw[draw=black,fill=color0] (axis cs:0.6,0) rectangle (axis cs:1.4,423.217412462928);
\draw[draw=black,fill=color0] (axis cs:1.6,0) rectangle (axis cs:2.4,443.320618047619);
\draw[draw=black,fill=color1] (axis cs:2.6,0) rectangle (axis cs:3.4,90.0043886682843);
\draw[draw=black,fill=color0] (axis cs:-0.4,443.457580821169) rectangle (axis cs:0.4,480.557580821169);
\draw[draw=black,fill=color0] (axis cs:0.6,423.217412462928) rectangle (axis cs:1.4,495.117412462928);
\draw[draw=black,fill=color0] (axis cs:1.6,443.320618047619) rectangle (axis cs:2.4,494.220618047619);
\draw[draw=black,fill=color1] (axis cs:2.6,90.0043886682843) rectangle (axis cs:3.4,130.104388668284);
\path [draw=white!50.1960784313725!black, semithick]
(axis cs:0,423.642269846524)
--(axis cs:0,463.272891795814);

\path [draw=white!50.1960784313725!black, semithick]
(axis cs:1,405.335647240982)
--(axis cs:1,441.099177684874);

\path [draw=white!50.1960784313725!black, semithick]
(axis cs:2,425.069015313276)
--(axis cs:2,461.572220781962);

\path [draw=white!50.1960784313725!black, semithick]
(axis cs:3,60.3708070310662)
--(axis cs:3,119.637970305502);

\path [draw=black, semithick]
(axis cs:0,467.660154721361)
--(axis cs:0,493.455006920977);

\path [draw=black, semithick]
(axis cs:1,479.33464971365)
--(axis cs:1,510.900175212206);

\path [draw=black, semithick]
(axis cs:2,479.903928486131)
--(axis cs:2,508.537307609106);

\path [draw=black, semithick]
(axis cs:3,119.737151813654)
--(axis cs:3,140.471625522914);

\addplot [semithick, white!50.1960784313725!black, mark=-, mark size=5, mark options={solid}, only marks]
table {%
0 423.642269846524
1 405.335647240982
2 425.069015313276
3 60.3708070310662
};
\addplot [semithick, white!50.1960784313725!black, mark=-, mark size=5, mark options={solid}, only marks]
table {%
0 463.272891795814
1 441.099177684874
2 461.572220781962
3 119.637970305502
};
\addplot [semithick, black, mark=-, mark size=5, mark options={solid}, only marks]
table {%
0 467.660154721361
1 479.33464971365
2 479.903928486131
3 119.737151813654
};
\addplot [semithick, black, mark=-, mark size=5, mark options={solid}, only marks]
table {%
0 493.455006920977
1 510.900175212206
2 508.537307609106
3 140.471625522914
};
\addplot [semithick, white!50.1960784313725!black, dashed]
table {%
-0.59 350
3.59 350
};
\addplot [semithick, black, dashed]
table {%
-0.59 400
3.59 400
};
\end{axis}

\end{tikzpicture}

%% file: Graphs/bvsf.1.tex
\begin{tikzpicture}

\definecolor{color0}{rgb}{0.529411764705882,0.807843137254902,0.92156862745098}
\definecolor{color1}{rgb}{0.913725490196078,0.588235294117647,0.47843137254902}
\definecolor{color2}{rgb}{0.596078431372549,0.984313725490196,0.596078431372549}

\begin{axis}[
height=\figH,
tick align=outside,
tick pos=left,
width=\figW,
x grid style={white!69.0196078431373!black},
xlabel={Agent},
xmin=-0.59, xmax=3.59,
xtick style={color=black},
xtick={0,1,2,3},
xticklabels={f,b,n0,n1},
y grid style={white!69.0196078431373!black},
ylabel={Mean cash+payout},
ymin=0, ymax=700,
ytick style={color=black}
]
\draw[draw=black,fill=color0] (axis cs:-0.4,0) rectangle (axis cs:0.4,536.966735709364);
\draw[draw=black,fill=color1] (axis cs:0.6,0) rectangle (axis cs:1.4,509.258656612324);
\draw[draw=black,fill=color2] (axis cs:1.6,0) rectangle (axis cs:2.4,222.073983865846);
\draw[draw=black,fill=color2] (axis cs:2.6,0) rectangle (axis cs:3.4,131.700623812466);
\draw[draw=black,fill=color0] (axis cs:-0.4,536.966735709364) rectangle (axis cs:0.4,634.766735709364);
\draw[draw=black,fill=color1] (axis cs:0.6,509.258656612324) rectangle (axis cs:1.4,545.258656612324);
\draw[draw=black,fill=color2] (axis cs:1.6,222.073983865846) rectangle (axis cs:2.4,254.273983865846);
\draw[draw=black,fill=color2] (axis cs:2.6,131.700623812466) rectangle (axis cs:3.4,165.700623812466);
\path [draw=white!50.1960784313725!black, semithick]
(axis cs:0,516.656651456357)
--(axis cs:0,557.276819962372);

\path [draw=white!50.1960784313725!black, semithick]
(axis cs:1,487.889593184226)
--(axis cs:1,530.627720040422);

\path [draw=white!50.1960784313725!black, semithick]
(axis cs:2,146.791933953147)
--(axis cs:2,297.356033778545);

\path [draw=white!50.1960784313725!black, semithick]
(axis cs:3,53.1502102038685)
--(axis cs:3,210.251037421063);

\path [draw=black, semithick]
(axis cs:0,618.673059932836)
--(axis cs:0,650.860411485892);

\path [draw=black, semithick]
(axis cs:1,534.939741940712)
--(axis cs:1,555.577571283935);

\path [draw=black, semithick]
(axis cs:2,243.583502118048)
--(axis cs:2,264.964465613644);

\path [draw=black, semithick]
(axis cs:3,155.091194044566)
--(axis cs:3,176.310053580366);

\addplot [semithick, white!50.1960784313725!black, mark=-, mark size=5, mark options={solid}, only marks]
table {%
0 516.656651456357
1 487.889593184226
2 146.791933953147
3 53.1502102038685
};
\addplot [semithick, white!50.1960784313725!black, mark=-, mark size=5, mark options={solid}, only marks]
table {%
0 557.276819962372
1 530.627720040422
2 297.356033778545
3 210.251037421063
};
\addplot [semithick, black, mark=-, mark size=5, mark options={solid}, only marks]
table {%
0 618.673059932836
1 534.939741940712
2 243.583502118048
3 155.091194044566
};
\addplot [semithick, black, mark=-, mark size=5, mark options={solid}, only marks]
table {%
0 650.860411485892
1 555.577571283935
2 264.964465613644
3 176.310053580366
};
\addplot [semithick, white!50.1960784313725!black, dashed]
table {%
-0.59 350
3.59 350
};
\addplot [semithick, black, dashed]
table {%
-0.59 400
3.59 400
};
\end{axis}

\end{tikzpicture}

%% file: Graphs/bvsf.2.tex
\begin{tikzpicture}

\definecolor{color0}{rgb}{0.529411764705882,0.807843137254902,0.92156862745098}
\definecolor{color1}{rgb}{0.913725490196078,0.588235294117647,0.47843137254902}
\definecolor{color2}{rgb}{0.596078431372549,0.984313725490196,0.596078431372549}

\begin{axis}[
height=\figH,
tick align=outside,
tick pos=left,
width=\figW,
x grid style={white!69.0196078431373!black},
xlabel={Agent},
xmin=-0.59, xmax=3.59,
xtick style={color=black},
xtick={0,1,2,3},
xticklabels={f0,f1,b,n},
y grid style={white!69.0196078431373!black},
ymin=0, ymax=700,
ytick style={color=black}
]
\draw[draw=black,fill=color0] (axis cs:-0.4,0) rectangle (axis cs:0.4,441.014798135273);
\draw[draw=black,fill=color0] (axis cs:0.6,0) rectangle (axis cs:1.4,430.104240182497);
\draw[draw=black,fill=color1] (axis cs:1.6,0) rectangle (axis cs:2.4,414.309258660814);
\draw[draw=black,fill=color2] (axis cs:2.6,0) rectangle (axis cs:3.4,114.571703021416);
\draw[draw=black,fill=color0] (axis cs:-0.4,441.014798135273) rectangle (axis cs:0.4,501.014798135273);
\draw[draw=black,fill=color0] (axis cs:0.6,430.104240182497) rectangle (axis cs:1.4,508.704240182497);
\draw[draw=black,fill=color1] (axis cs:1.6,414.309258660814) rectangle (axis cs:2.4,451.709258660814);
\draw[draw=black,fill=color2] (axis cs:2.6,114.571703021416) rectangle (axis cs:3.4,138.571703021416);
\path [draw=white!50.1960784313725!black, semithick]
(axis cs:0,425.616554503007)
--(axis cs:0,456.413041767539);

\path [draw=white!50.1960784313725!black, semithick]
(axis cs:1,415.650478326411)
--(axis cs:1,444.558002038583);

\path [draw=white!50.1960784313725!black, semithick]
(axis cs:2,389.133050621811)
--(axis cs:2,439.485466699816);

\path [draw=white!50.1960784313725!black, semithick]
(axis cs:3,81.1087694156988)
--(axis cs:3,148.034636627133);

\path [draw=black, semithick]
(axis cs:0,485.634290991489)
--(axis cs:0,516.395305279057);

\path [draw=black, semithick]
(axis cs:1,492.43151453496)
--(axis cs:1,524.976965830034);

\path [draw=black, semithick]
(axis cs:2,440.92920300233)
--(axis cs:2,462.489314319297);

\path [draw=black, semithick]
(axis cs:3,130.413271799667)
--(axis cs:3,146.730134243164);

\addplot [semithick, white!50.1960784313725!black, mark=-, mark size=5, mark options={solid}, only marks]
table {%
0 425.616554503007
1 415.650478326411
2 389.133050621811
3 81.1087694156988
};
\addplot [semithick, white!50.1960784313725!black, mark=-, mark size=5, mark options={solid}, only marks]
table {%
0 456.413041767539
1 444.558002038583
2 439.485466699816
3 148.034636627133
};
\addplot [semithick, black, mark=-, mark size=5, mark options={solid}, only marks]
table {%
0 485.634290991489
1 492.43151453496
2 440.92920300233
3 130.413271799667
};
\addplot [semithick, black, mark=-, mark size=5, mark options={solid}, only marks]
table {%
0 516.395305279057
1 524.976965830034
2 462.489314319297
3 146.730134243164
};
\addplot [semithick, white!50.1960784313725!black, dashed]
table {%
-0.59 350
3.59 350
};
\addplot [semithick, black, dashed]
table {%
-0.59 400
3.59 400
};
\end{axis}

\end{tikzpicture}

%% file: Graphs/bvsf.3.tex
\begin{tikzpicture}

\definecolor{color0}{rgb}{0.529411764705882,0.807843137254902,0.92156862745098}
\definecolor{color1}{rgb}{0.913725490196078,0.588235294117647,0.47843137254902}

\begin{axis}[
height=\figH,
tick align=outside,
tick pos=left,
width=\figW,
x grid style={white!69.0196078431373!black},
xlabel={Agent},
xmin=-0.59, xmax=3.59,
xtick style={color=black},
xtick={0,1,2,3},
xticklabels={f0,f1,f2,b},
y grid style={white!69.0196078431373!black},
ymin=0, ymax=700,
ytick style={color=black}
]
\draw[draw=black,fill=color0] (axis cs:-0.4,0) rectangle (axis cs:0.4,378.247916343537);
\draw[draw=black,fill=color0] (axis cs:0.6,0) rectangle (axis cs:1.4,370.39560301737);
\draw[draw=black,fill=color0] (axis cs:1.6,0) rectangle (axis cs:2.4,375.032120573745);
\draw[draw=black,fill=color1] (axis cs:2.6,0) rectangle (axis cs:3.4,276.324360065348);
\draw[draw=black,fill=color0] (axis cs:-0.4,378.247916343537) rectangle (axis cs:0.4,434.047916343537);
\draw[draw=black,fill=color0] (axis cs:0.6,370.39560301737) rectangle (axis cs:1.4,424.09560301737);
\draw[draw=black,fill=color0] (axis cs:1.6,375.032120573745) rectangle (axis cs:2.4,426.932120573745);
\draw[draw=black,fill=color1] (axis cs:2.6,276.324360065348) rectangle (axis cs:3.4,314.924360065348);
\path [draw=white!50.1960784313725!black, semithick]
(axis cs:0,368.375349917897)
--(axis cs:0,388.120482769178);

\path [draw=white!50.1960784313725!black, semithick]
(axis cs:1,360.878177626713)
--(axis cs:1,379.913028408026);

\path [draw=white!50.1960784313725!black, semithick]
(axis cs:2,365.287779004403)
--(axis cs:2,384.776462143087);

\path [draw=white!50.1960784313725!black, semithick]
(axis cs:3,262.274296627858)
--(axis cs:3,290.374423502838);

\path [draw=black, semithick]
(axis cs:0,418.750387131275)
--(axis cs:0,449.345445555799);

\path [draw=black, semithick]
(axis cs:1,408.73941859545)
--(axis cs:1,439.451787439289);

\path [draw=black, semithick]
(axis cs:2,411.894314882754)
--(axis cs:2,441.969926264736);

\path [draw=black, semithick]
(axis cs:3,305.454323106603)
--(axis cs:3,324.394397024094);

\addplot [semithick, white!50.1960784313725!black, mark=-, mark size=5, mark options={solid}, only marks]
table {%
0 368.375349917897
1 360.878177626713
2 365.287779004403
3 262.274296627858
};
\addplot [semithick, white!50.1960784313725!black, mark=-, mark size=5, mark options={solid}, only marks]
table {%
0 388.120482769178
1 379.913028408026
2 384.776462143087
3 290.374423502838
};
\addplot [semithick, black, mark=-, mark size=5, mark options={solid}, only marks]
table {%
0 418.750387131275
1 408.73941859545
2 411.894314882754
3 305.454323106603
};
\addplot [semithick, black, mark=-, mark size=5, mark options={solid}, only marks]
table {%
0 449.345445555799
1 439.451787439289
2 441.969926264736
3 324.394397024094
};
\addplot [semithick, white!50.1960784313725!black, dashed]
table {%
-0.59 350
3.59 350
};
\addplot [semithick, black, dashed]
table {%
-0.59 400
3.59 400
};
\end{axis}

\end{tikzpicture}

%% file: Graphs/2bf_fun.1.tex
\begin{tikzpicture}

\definecolor{color0}{rgb}{0.529411764705882,0.807843137254902,0.92156862745098}
\definecolor{color1}{rgb}{0.913725490196078,0.588235294117647,0.47843137254902}
\definecolor{color2}{rgb}{0.596078431372549,0.984313725490196,0.596078431372549}

\begin{axis}[
height=\figH,
tick align=outside,
tick pos=left,
width=\figW,
x grid style={white!69.0196078431373!black},
xlabel={Agent},
xmin=-0.59, xmax=3.59,
xtick style={color=black},
xtick={0,1,2,3},
xticklabels={f,b0,b1,n},
y grid style={white!69.0196078431373!black},
ylabel={Mean cash+payout},
ymin=0, ymax=700,
ytick style={color=black}
]
\draw[draw=black,fill=color0] (axis cs:-0.4,0) rectangle (axis cs:0.4,520.728341083804);
\draw[draw=black,fill=color1] (axis cs:0.6,0) rectangle (axis cs:1.4,393.627473867896);
\draw[draw=black,fill=color1] (axis cs:1.6,0) rectangle (axis cs:2.4,394.083962491224);
\draw[draw=black,fill=color2] (axis cs:2.6,0) rectangle (axis cs:3.4,91.5602225570767);
\draw[draw=black,fill=color0] (axis cs:-0.4,520.728341083804) rectangle (axis cs:0.4,619.153341083804);
\draw[draw=black,fill=color1] (axis cs:0.6,393.627473867896) rectangle (axis cs:1.4,425.702473867896);
\draw[draw=black,fill=color1] (axis cs:1.6,394.083962491224) rectangle (axis cs:2.4,427.008962491224);
\draw[draw=black,fill=color2] (axis cs:2.6,91.5602225570767) rectangle (axis cs:3.4,128.135222557077);
\path [draw=white!50.1960784313725!black, semithick]
(axis cs:0,507.446314082695)
--(axis cs:0,534.010368084912);

\path [draw=white!50.1960784313725!black, semithick]
(axis cs:1,384.442928995807)
--(axis cs:1,402.812018739985);

\path [draw=white!50.1960784313725!black, semithick]
(axis cs:2,384.021042411527)
--(axis cs:2,404.146882570922);

\path [draw=white!50.1960784313725!black, semithick]
(axis cs:3,71.3751018848742)
--(axis cs:3,111.745343229279);

\path [draw=black, semithick]
(axis cs:0,610.983720912094)
--(axis cs:0,627.322961255513);

\path [draw=black, semithick]
(axis cs:1,420.522759228072)
--(axis cs:1,430.882188507719);

\path [draw=black, semithick]
(axis cs:2,421.769502237534)
--(axis cs:2,432.248422744915);

\path [draw=black, semithick]
(axis cs:3,122.29611357111)
--(axis cs:3,133.974331543044);

\addplot [semithick, white!50.1960784313725!black, mark=-, mark size=5, mark options={solid}, only marks]
table {%
0 507.446314082695
1 384.442928995807
2 384.021042411527
3 71.3751018848742
};
\addplot [semithick, white!50.1960784313725!black, mark=-, mark size=5, mark options={solid}, only marks]
table {%
0 534.010368084912
1 402.812018739985
2 404.146882570922
3 111.745343229279
};
\addplot [semithick, black, mark=-, mark size=5, mark options={solid}, only marks]
table {%
0 610.983720912094
1 420.522759228072
2 421.769502237534
3 122.29611357111
};
\addplot [semithick, black, mark=-, mark size=5, mark options={solid}, only marks]
table {%
0 627.322961255513
1 430.882188507719
2 432.248422744915
3 133.974331543044
};
\addplot [semithick, white!50.1960784313725!black, dashed]
table {%
-0.59 350
3.59 350
};
\addplot [semithick, black, dashed]
table {%
-0.59 400
3.59 400
};
\end{axis}

\end{tikzpicture}

%% file: Graphs/2bf_fun.2.tex
\begin{tikzpicture}

\definecolor{color0}{rgb}{0.529411764705882,0.807843137254902,0.92156862745098}
\definecolor{color1}{rgb}{0.913725490196078,0.588235294117647,0.47843137254902}

\begin{axis}[
height=\figH,
tick align=outside,
tick pos=left,
width=\figW,
x grid style={white!69.0196078431373!black},
xlabel={Agent},
xmin=-0.59, xmax=3.59,
xtick style={color=black},
xtick={0,1,2,3},
xticklabels={f0,f1,b0,b1},
y grid style={white!69.0196078431373!black},
ymin=0, ymax=700,
ytick style={color=black}
]
\draw[draw=black,fill=color0] (axis cs:-0.4,0) rectangle (axis cs:0.4,415.921150824557);
\draw[draw=black,fill=color0] (axis cs:0.6,0) rectangle (axis cs:1.4,430.289247125803);
\draw[draw=black,fill=color1] (axis cs:1.6,0) rectangle (axis cs:2.4,282.563711706596);
\draw[draw=black,fill=color1] (axis cs:2.6,0) rectangle (axis cs:3.4,271.225890343045);
\draw[draw=black,fill=color0] (axis cs:-0.4,415.921150824557) rectangle (axis cs:0.4,489.921150824557);
\draw[draw=black,fill=color0] (axis cs:0.6,430.289247125803) rectangle (axis cs:1.4,492.714247125803);
\draw[draw=black,fill=color1] (axis cs:1.6,282.563711706596) rectangle (axis cs:2.4,314.013711706596);
\draw[draw=black,fill=color1] (axis cs:2.6,271.225890343045) rectangle (axis cs:3.4,303.350890343045);
\path [draw=white!50.1960784313725!black, semithick]
(axis cs:0,407.478608526805)
--(axis cs:0,424.363693122308);

\path [draw=white!50.1960784313725!black, semithick]
(axis cs:1,421.067579227965)
--(axis cs:1,439.510915023641);

\path [draw=white!50.1960784313725!black, semithick]
(axis cs:2,271.918671943656)
--(axis cs:2,293.208751469535);

\path [draw=white!50.1960784313725!black, semithick]
(axis cs:3,260.331860518319)
--(axis cs:3,282.119920167771);

\path [draw=black, semithick]
(axis cs:0,481.736408922701)
--(axis cs:0,498.105892726413);

\path [draw=black, semithick]
(axis cs:1,484.695912276846)
--(axis cs:1,500.73258197476);

\path [draw=black, semithick]
(axis cs:2,309.295240026062)
--(axis cs:2,318.73218338713);

\path [draw=black, semithick]
(axis cs:3,298.637362603655)
--(axis cs:3,308.064418082435);

\addplot [semithick, white!50.1960784313725!black, mark=-, mark size=5, mark options={solid}, only marks]
table {%
0 407.478608526805
1 421.067579227965
2 271.918671943656
3 260.331860518319
};
\addplot [semithick, white!50.1960784313725!black, mark=-, mark size=5, mark options={solid}, only marks]
table {%
0 424.363693122308
1 439.510915023641
2 293.208751469535
3 282.119920167771
};
\addplot [semithick, black, mark=-, mark size=5, mark options={solid}, only marks]
table {%
0 481.736408922701
1 484.695912276846
2 309.295240026062
3 298.637362603655
};
\addplot [semithick, black, mark=-, mark size=5, mark options={solid}, only marks]
table {%
0 498.105892726413
1 500.73258197476
2 318.73218338713
3 308.064418082435
};
\addplot [semithick, white!50.1960784313725!black, dashed]
table {%
-0.59 350
3.59 350
};
\addplot [semithick, black, dashed]
table {%
-0.59 400
3.59 400
};
\end{axis}

\end{tikzpicture}

%% file: Graphs/speeds.tex
\begin{tikzpicture}

\definecolor{color0}{rgb}{0.529411764705882,0.807843137254902,0.92156862745098}

\begin{axis}[
height=\figH,
tick align=outside,
tick pos=left,
width=\figW,
x grid style={white!69.0196078431373!black},
xlabel={Agent},
xmin=-0.59, xmax=3.59,
xtick style={color=black},
xtick={0,1,2,3},
xticklabels={fast,slow0,slow1,slow2},
y grid style={white!69.0196078431373!black},
ylabel={Mean cash+payout},
ymin=0, ymax=700,
ytick style={color=black}
]
\draw[draw=black,fill=color0] (axis cs:-0.4,0) rectangle (axis cs:0.4,303.890487351668);
\draw[draw=black,fill=color0] (axis cs:0.6,0) rectangle (axis cs:1.4,365.769941408004);
\draw[draw=black,fill=color0] (axis cs:1.6,0) rectangle (axis cs:2.4,364.664793941895);
\draw[draw=black,fill=color0] (axis cs:2.6,0) rectangle (axis cs:3.4,365.674777298433);
\draw[draw=black,fill=color0] (axis cs:-0.4,303.890487351668) rectangle (axis cs:0.4,380.990487351668);
\draw[draw=black,fill=color0] (axis cs:0.6,365.769941408004) rectangle (axis cs:1.4,405.794941408004);
\draw[draw=black,fill=color0] (axis cs:1.6,364.664793941895) rectangle (axis cs:2.4,403.814793941895);
\draw[draw=black,fill=color0] (axis cs:2.6,365.674777298433) rectangle (axis cs:3.4,409.399777298433);
\path [draw=white!50.1960784313725!black, semithick]
(axis cs:0,301.410942210574)
--(axis cs:0,306.370032492761);

\path [draw=white!50.1960784313725!black, semithick]
(axis cs:1,363.493957215721)
--(axis cs:1,368.045925600287);

\path [draw=white!50.1960784313725!black, semithick]
(axis cs:2,362.44232112074)
--(axis cs:2,366.887266763051);

\path [draw=white!50.1960784313725!black, semithick]
(axis cs:3,363.551597513044)
--(axis cs:3,367.797957083821);

\path [draw=black, semithick]
(axis cs:0,372.683809315292)
--(axis cs:0,389.297165388043);

\path [draw=black, semithick]
(axis cs:1,400.037121182433)
--(axis cs:1,411.552761633576);

\path [draw=black, semithick]
(axis cs:2,398.210976046924)
--(axis cs:2,409.418611836867);

\path [draw=black, semithick]
(axis cs:3,403.464817841958)
--(axis cs:3,415.334736754907);

\addplot [semithick, white!50.1960784313725!black, mark=-, mark size=5, mark options={solid}, only marks]
table {%
0 301.410942210574
1 363.493957215721
2 362.44232112074
3 363.551597513044
};
\addplot [semithick, white!50.1960784313725!black, mark=-, mark size=5, mark options={solid}, only marks]
table {%
0 306.370032492761
1 368.045925600287
2 366.887266763051
3 367.797957083821
};
\addplot [semithick, black, mark=-, mark size=5, mark options={solid}, only marks]
table {%
0 372.683809315292
1 400.037121182433
2 398.210976046924
3 403.464817841958
};
\addplot [semithick, black, mark=-, mark size=5, mark options={solid}, only marks]
table {%
0 389.297165388043
1 411.552761633576
2 409.418611836867
3 415.334736754907
};
\addplot [semithick, white!50.1960784313725!black, dashed]
table {%
-0.59 350
3.59 350
};
\addplot [semithick, black, dashed]
table {%
-0.59 400
3.59 400
};
\end{axis}

\end{tikzpicture}

%% file: Graphs/acf/acf.50.tex
\begin{tikzpicture}

\begin{axis}[
height=\figH,
tick align=outside,
tick pos=left,
title={\(\displaystyle \Delta t=50\)},
width=\figW,
x grid style={white!69.0196078431373!black},
xlabel={Lag},
xmin=-1, xmax=21,
xtick style={color=black},
y grid style={white!69.0196078431373!black},
ymin=-0.6, ymax=1,
ytick style={color=black}
]
\addplot [semithick, black, opacity=0.1]
table {%
0 1
1 0.098332014524441
2 -0.0525530137251019
3 -0.0735419176227968
4 0.0299120678545809
5 0.112620044491841
6 -0.0122099296726972
7 -0.0156198712386648
8 -0.0351254854093173
9 0.00175459336754005
10 -0.0338853533080002
11 0.00567700841956852
12 -0.209443192336773
13 0.0601499587630213
14 0.118522476582032
15 -0.112851918775328
16 -0.0603461741787168
17 -0.176169661066395
18 -0.0668342607911356
19 0.0403170032803606
20 0.0623700921088689
};
\addplot [semithick, black, opacity=0.1]
table {%
0 1
1 0.0747145095310001
2 0.313473276619193
3 0.0131434821173889
4 -0.0701707746455794
5 -0.0370278145525324
6 -0.0336675672504489
7 0.13576411529671
8 0.0417162222365468
9 0.0199978564567222
10 0.0162961709584357
11 -0.0882703644600339
12 0.0307978288362163
13 -0.118778849921035
14 0.0686285640890279
15 -0.0438566352732746
16 0.0474057068834541
17 0.0495093340754478
18 -0.118529225927436
19 0.0957352510682699
20 -0.0662052909809384
};
\addplot [semithick, black, opacity=0.1]
table {%
0 1
1 -0.286136901281995
2 0.129225836335108
3 0.0236607662305169
4 -0.103782109647584
5 0.264940346067482
6 0.0246445843671013
7 0.0289843080801462
8 0.0119439227719887
9 -0.261282462540236
10 0.0592280351763805
11 0.087234990431652
12 -0.0439731951858534
13 0.083330305891674
14 -0.317683309259756
15 0.099345901366224
16 0.0715443955499255
17 -0.103852259629967
18 0.238833006388403
19 -0.177702291280716
20 -0.107008143784606
};
\addplot [semithick, black, opacity=0.1]
table {%
0 1
1 -0.141477543919664
2 0.0833259281380483
3 0.108590515542467
4 -0.256821876093948
5 -0.11585360186983
6 -0.148674823500925
7 -0.0596781703603027
8 0.0275082066623344
9 -0.127896267490969
10 0.0993835961691676
11 -0.0262840113685648
12 -0.0822716189143616
13 0.227008105058267
14 -0.113599230197013
15 0.154345325676608
16 -0.00948497871718209
17 -0.149483709204813
18 -0.0394423276537385
19 -0.0225835643126233
20 -0.0895687994489349
};
\addplot [semithick, black, opacity=0.1]
table {%
0 1
1 -0.0383236178479343
2 -0.226848493490706
3 -0.0375399800105877
4 -0.152624665879757
5 -0.0734364429541372
6 0.343315799382995
7 -0.0285057750126844
8 -0.0546742283086663
9 0.0994168806521031
10 0.0956314291314218
11 -0.128429319407993
12 -0.102989056353403
13 -0.106520020678225
14 -0.039401512993276
15 -0.00689966382169496
16 0.063424773041154
17 0.0910712022901775
18 -0.10211165119768
19 0.0616502200595856
20 -0.134504479812688
};
\addplot [semithick, black, opacity=0.1]
table {%
0 1
1 -0.380750441516126
2 -0.0361655287377426
3 -0.0845647742383579
4 0.149399788884307
5 -0.0365410912934196
6 -0.130459237527419
7 0.0701885784004188
8 0.112285380292975
9 -0.0865744646508714
10 -0.0152375894253431
11 0.128238740657909
12 -0.107476198732409
13 -0.05810299287418
14 0.0246242716078668
15 0.15086833079219
16 -0.155034644937438
17 0.0984414088437276
18 0.0119766750419511
19 0.0168358147800207
20 -0.0418000308552041
};
\addplot [semithick, black, opacity=0.1]
table {%
0 1
1 -0.257037918286632
2 -0.0904002970461809
3 0.0325939685327049
4 0.0904260732223436
5 -0.139287859074065
6 -0.0887125464827166
7 -0.18485236728509
8 0.036709324563604
9 0.124214443578675
10 -0.0191926077096558
11 -0.164506986676439
12 0.306423307724707
13 -0.0601580562304127
14 -0.0286903518642219
15 -0.032468975266421
16 -0.0485918603748458
17 0.0405998086392452
18 0.053114670209045
19 -0.0905295545303714
20 -0.00445735770923722
};
\addplot [semithick, black, opacity=0.1]
table {%
0 1
1 0.34586827208017
2 0.117491577877953
3 -0.219052095330395
4 -0.176289924846678
5 -0.305125296631515
6 -0.0256042257259954
7 -0.127527204111628
8 0.0506652335024611
9 -0.0420793701480967
10 -0.0274538944759661
11 -0.00433421778233464
12 0.157984688158765
13 0.0581748915995753
14 -0.0301978124188325
15 -0.077127470762316
16 -0.0600184728396548
17 -0.00673132149636028
18 0.0279677669648349
19 -0.0486995307098159
20 -0.0185727025218811
};
\addplot [semithick, black, opacity=0.1]
table {%
0 1
1 -0.226655109600936
2 0.0361421931218402
3 -0.145986887054649
4 -0.104669224139427
5 0.0690583981227424
6 -0.0554394719920366
7 -0.0351001136390571
8 0.164071433915929
9 -0.0712661897206648
10 0.119185127964021
11 -0.161256662380573
12 -0.00242607893600577
13 -0.0386641224740572
14 -0.0934340337974334
15 -0.0838042533230342
16 0.0958869792871339
17 -0.0450089048183573
18 0.157221289320208
19 -0.0658889545144148
20 -0.037914731633332
};
\addplot [semithick, black, opacity=0.1]
table {%
0 1
1 -0.296243984972488
2 0.191007785314458
3 -0.18268879104984
4 0.0137914928988943
5 0.000408201074217947
6 -0.161187984909694
7 0.0605614182865197
8 0.0293087305124469
9 0.018773089660509
10 -0.0462447874299668
11 0.0215410667923394
12 -0.0868283097724358
13 -0.0768825763051512
14 -0.0467025260017268
15 0.0263407562317811
16 0.200553388032865
17 -0.187334158321743
18 0.0978983829718632
19 -0.0573356977439783
20 -0.00407269913328235
};
\addplot [semithick, black, opacity=0.1]
table {%
0 1
1 -0.257200270586779
2 0.138613541165621
3 0.0208540869502651
4 -0.0353884029276613
5 -0.0454704721680573
6 -0.0136976917863661
7 0.0489853563167753
8 -0.147484177012479
9 0.226289521303023
10 -0.0737845473847282
11 0.0596441369997673
12 -0.016704470406431
13 -0.0348130153200169
14 0.0201683470403029
15 0.0186806627141386
16 0.00653931971741616
17 -0.0103551391499244
18 0.0314872451973935
19 -0.0127197011638394
20 -0.014047652200686
};
\addplot [semithick, black, opacity=0.1]
table {%
0 1
1 -0.231663819038047
2 0.0946844567504194
3 -0.0615932338582996
4 -0.145356083029701
5 0.12609024512966
6 0.0416580260932098
7 0.0554326770319956
8 -0.0201715888565718
9 -0.0356849854826549
10 -0.0841899994422497
11 0.0158820293645963
12 0.146729430254064
13 -0.105234866173636
14 -0.0165031516050489
15 -0.0820869549168233
16 -0.109994577282266
17 0.292529449091783
18 -0.0213923119521396
19 0.128385133517828
20 -0.136478419273219
};
\addplot [semithick, black, opacity=0.1]
table {%
0 1
1 -0.211720779145796
2 -0.0229996872155935
3 0.133325536192334
4 -0.0998776371081404
5 0.0210438568791687
6 -0.00378532875159958
7 -0.0433865475053366
8 0.135707827509503
9 -0.126984077955684
10 -0.0344701114816616
11 -0.0942484568731129
12 -0.12212346508718
13 -0.00710264748963903
14 -0.0036058077841065
15 0.0998236263093027
16 -0.0122141580458349
17 0.0192094823742329
18 -0.020067494063829
19 -0.0908858484210032
20 0.0940682320373369
};
\addplot [semithick, black, opacity=0.1]
table {%
0 1
1 -0.0148074699070529
2 -0.00715778326722453
3 -0.0151589051838359
4 -0.0188098291442593
5 0.000808476725068263
6 -0.0328847775031324
7 0.146405434117769
8 -0.047962392799751
9 0.0283669910960302
10 -0.0881541404245553
11 -0.0302724537498612
12 -0.0341388901116828
13 -0.0477740277463692
14 -0.0112412654441581
15 -0.0334459252001845
16 -0.00150999086290941
17 -0.0418136486198091
18 -0.0301243516510284
19 0.0270850114064648
20 -0.0442243786458679
};
\addplot [semithick, black, opacity=0.1]
table {%
0 1
1 -0.186594085147587
2 -0.0765855962995154
3 0.0856674413889637
4 0.0285882707476942
5 0.0794586908691521
6 -0.100721471877124
7 -0.121859213213723
8 0.14239341387244
9 -0.0162675108399768
10 -0.182270110313582
11 -0.0131445997637289
12 -0.0286775302884986
13 -0.147876550240995
14 0.15255609058637
15 -0.178189045392177
16 0.112016965580754
17 0.0863021650226061
18 -0.0795025693685579
19 -0.0566115312234296
20 0.0364415904530293
};
\addplot [semithick, black, opacity=0.1]
table {%
0 1
1 -0.13143732593689
2 0.0531513941504877
3 -0.0291446011377975
4 -0.143270368346032
5 0.119661333665168
6 0.0600029640138683
7 0.0181713425180006
8 -0.188938163390636
9 -0.0901784788735526
10 0.0242098356679116
11 0.0563701759186463
12 0.0606959139297022
13 0.0759888114240708
14 -0.134222559314987
15 0.179850756214695
16 -0.120006706889615
17 0.112744291219313
18 -0.02086099557723
19 0.000750695739609527
20 -0.00618662878377504
};
\addplot [semithick, black, opacity=0.1]
table {%
0 1
1 -0.206415846819607
2 -0.0465174589882985
3 -0.0462060691224499
4 -0.0886113446242868
5 0.0542684885444864
6 -0.0766328315120844
7 0.0398698121098183
8 -0.122508866019809
9 -0.0287615892482129
10 -0.0662578406799996
11 0.387336891887796
12 -0.105114820793232
13 -0.100467742458951
14 -0.170665555221144
15 0.0497312787520559
16 0.0832446771061549
17 -0.161117547478326
18 0.0849474532184042
19 -0.0243519900173658
20 0.0551535217947733
};
\addplot [semithick, black, opacity=0.1]
table {%
0 1
1 -0.293222603618079
2 0.191592379047187
3 0.222549294706
4 -0.149838986250448
5 0.181248866996478
6 -0.160824625196487
7 0.0334700666319998
8 -0.00165182156756406
9 0.00789376732321019
10 -0.0906266894283476
11 0.108767298737769
12 -0.154318768325083
13 0.0819718873926878
14 -0.0193967647637338
15 -0.0862073000782146
16 0.166904118342511
17 -0.161043855942849
18 0.0417444095746253
19 0.0738730066975179
20 -0.144891160102801
};
\addplot [semithick, black, opacity=0.1]
table {%
0 1
1 -0.292408772007768
2 -0.119995680958054
3 0.277000292024043
4 -0.159904431942052
5 0.0086431213189865
6 -0.213334528435155
};
\addplot [semithick, black, opacity=0.1]
table {%
0 1
1 -0.0412318154118351
2 -0.395356736978019
3 -0.0100214605870875
4 0.219350450954159
5 -0.00351832877551441
6 0.086252058351923
7 -0.0226022794631507
8 -0.105967332256737
9 0.0966397952880076
10 0.0411408057805829
11 -0.186715237995345
12 0.0227323745328845
13 0.217078255639374
14 -0.075516647315421
15 -0.0809901863890462
16 0.0917770277207895
17 -0.00269908833401923
18 -0.17852845032336
19 -0.00715758312209279
20 0.220193990085813
};
\addplot [semithick, black, opacity=0.1]
table {%
0 1
1 -0.0538821062304539
2 0.0317222333828656
3 0.3124483517698
4 -0.00982687111459389
5 0.00224699814455445
6 -0.00514719242137162
7 -0.019294703177025
8 -0.0421570771509238
9 -0.0424007171695783
10 -0.165163723147509
11 -0.0586747772534362
12 -0.0201102573100016
13 -0.0515906779786652
14 -0.0113109619735574
15 -0.00591079445654716
16 -0.0234694968342786
17 0.0808027565187746
18 0.00438727625884429
19 -0.0022526011853193
20 0.0463590949139541
};
\addplot [semithick, black, opacity=0.1]
table {%
0 1
1 -0.177200749985712
2 -0.0506332477280199
3 0.00367131704771608
4 0.0261971255011031
5 0.0252901909587349
6 -0.174698416925261
7 -0.00892049953958196
8 0.00931473120172804
9 -0.0913912522346038
10 0.173821105473222
11 -0.0835698749733118
12 -0.0374912179313851
13 -0.056593841628331
14 -0.0984901204371051
15 0.032945888051643
16 -0.0542556418539466
17 -0.0109069289044168
18 0.0708902552503071
19 0.0163174220035762
20 0.00617067899782212
};
\addplot [semithick, black, opacity=0.1]
table {%
0 1
1 0.0732555336645716
2 -0.0334275294291143
3 -0.0181276664478343
4 -0.0743698658567041
5 0.00276938551141426
6 0.0546650892582598
7 -0.0407658577272779
8 -0.0448181957452923
9 -0.00377820874243282
10 0.0214043410295416
11 0.0217365396870826
12 0.00540495513620025
13 -0.0326571259755878
14 -0.0681252045756284
15 -0.00366589998194628
16 0.0158367748286408
17 0.00269970454369461
18 -0.0801938916056094
19 -0.00422747498137774
20 0.0803421761182832
};
\addplot [semithick, black, opacity=0.1]
table {%
0 1
1 0.267621997486326
2 0.1572875134241
3 -0.0901858038798761
4 -0.318917924206437
5 0.0807877749392599
6 -0.0817329081492971
7 -0.0144152163894717
8 -0.0265463155033053
9 -0.110357179926009
10 0.0952359057955538
11 0.0615460319401951
12 -0.0442757651854277
13 0.0104543482670485
14 -0.0595495442671293
15 -0.00819181762979109
16 0.0326601558709492
17 0.0067794853995892
18 0.011371016265583
19 -0.0269839812204296
20 -0.00429970179250547
};
\addplot [semithick, black, opacity=0.1]
table {%
0 1
1 -0.109620927186897
2 0.0866702810426222
3 -0.156361005173322
4 0.00241650265518938
5 0.0336545076210191
6 0.0836072174053294
7 -0.184589863665112
8 0.0504415686294355
9 0.0706028170416503
10 0.0445334074384253
11 -0.00245693324910168
12 0.00148497168446631
13 0.0119349210858444
14 -0.00292759622225275
15 -0.0316106339143927
16 -0.0294680272594787
17 -0.0224183033490129
18 0.00884566601440235
19 0.00189398114656231
20 0.0265148543178165
};
\addplot [semithick, black, opacity=0.1]
table {%
0 1
1 0.375942046362854
2 0.130738658854162
3 0.183261088418273
4 0.116161671397575
5 0.0336793486951595
6 0.125955144175883
7 0.135173651084094
8 0.115473411770513
9 -0.0524632762520842
10 -0.107418997003672
11 -0.0345489867377984
12 -0.0604082404419128
13 -0.0323985450817751
14 -0.19478830232192
15 0.0397080462919304
16 0.0499635397857141
17 -0.0509880894703387
18 -0.0215213923070805
19 0.101526890036723
20 -0.0781264383184728
};
\addplot [semithick, black, opacity=0.1]
table {%
0 1
1 0.305409782701097
2 -0.0410679562165374
3 0.0360899905594617
4 0.145699257252816
5 0.225420585669994
6 -0.0141342669425898
7 -0.100285770463852
8 -0.0631271862634049
9 0.0490701021747393
10 0.0117253107074882
11 -0.0946268603438699
12 -0.0172553849044444
13 -0.0531487264195782
14 -0.0366453660117359
15 -0.0241084396842756
16 0.0146224433256928
17 0.0214757669046548
18 -0.10667958527021
19 -0.022588006920084
20 0.00360284815489294
};
\addplot [semithick, black, opacity=0.1]
table {%
0 1
1 0.0196933028622443
2 -0.0638350616687955
3 -0.152179249493781
4 0.190121152948232
5 0.101271928361513
6 -0.115021922317849
7 -0.195857443980633
8 0.0863420245544439
9 -0.0454373926016151
10 0.0148799412891058
11 0.0224054759214509
12 -0.0124510103344825
13 0.0352647769066612
14 -0.0893028122124784
15 -0.0766862443991519
16 -0.201198997152241
17 -0.0771725341021641
18 -0.0242233074683602
19 -0.0446734437473624
20 -0.046720216347847
};
\addplot [semithick, black, opacity=0.1]
table {%
0 1
1 -0.335897629088373
2 -0.185876778405192
3 0.258565525485678
4 -0.0687902575892015
5 -0.235535188053622
6 0.246204522643913
7 0.0193923838352673
8 -0.244160770161343
9 0.266438028900516
10 0.0299318753063306
11 -0.150850642302594
12 0.037130689128643
13 0.0454864981171564
14 -0.141177091121968
15 -0.0048618410921178
16 0.0251219971532715
17 -0.0214926326053513
18 -0.0442829338502171
19 0.135815225586621
20 -0.12723476186176
};
\addplot [semithick, black, opacity=0.1]
table {%
0 1
1 -0.176867888164876
2 -0.0269899774204551
3 -0.0218732689959526
4 0.0241549245968085
5 -0.0531598732408494
6 0.046842189902743
7 0.134148317605637
8 -0.175224029925497
9 0.0703903628196639
10 -0.0993813428498368
11 0.0930545975893368
12 -0.0237656714822305
13 0.0320255220823867
14 -0.0261999782228533
15 -0.0795316913228332
16 -2.94262308132671e-05
17 -0.0334261004243732
18 0.0776594362741824
19 -0.0707600256630621
20 0.0389462980162606
};
\addplot [semithick, black, opacity=0.1]
table {%
0 1
1 -0.298390142544577
2 0.193089904714835
3 -0.0902232699625208
4 0.048686999688679
5 -0.13259708639999
6 0.106017678304725
7 -0.252705605144521
8 0.225888403148675
9 -0.105644433638638
10 -0.0511735813011263
11 0.0217814676493337
12 0.0705658582066705
13 -0.239269856168762
14 0.182870279671017
15 -0.0312704374942033
16 0.0395930387203623
17 0.0209965820617745
18 -0.115426990124729
19 0.00255166197819587
20 0.0869345101858807
};
\addplot [semithick, black, opacity=0.1]
table {%
0 1
1 0.0302524356629383
2 -0.177775978478192
3 0.171599019106091
4 0.130469850765181
5 0.0301755111343138
6 -0.00442970360032001
7 0.156598867075009
8 0.0399789397749334
9 0.078820457765133
10 -0.0350870489976833
11 0.0113019218236465
12 0.0367543321338235
13 -0.0318224626211236
14 0.000135377227210556
15 -0.102681208235628
16 -0.118218994309419
17 -0.0467856622588392
18 0.094873371925255
19 -0.124940070402024
20 -0.178014494270483
};
\addplot [semithick, black, opacity=0.1]
table {%
0 1
1 -0.229140463528757
2 0.133502824821331
3 -0.128126819940461
4 0.0378234475115688
5 -0.10055677861798
6 0.0145934895766086
7 -0.0678098939703495
8 0.124093324326137
9 -0.203174828034345
10 0.115169335677474
11 -0.123295136019366
12 0.217647149474176
13 0.0238566170826674
14 0.0632117816032894
15 -0.00619957473343861
16 0.072614003997103
17 -0.0864074539670483
18 -0.0056771005241578
19 -0.0712342193794187
20 -0.0441389655769542
};
\addplot [semithick, black, opacity=0.1]
table {%
0 1
1 -0.128390244471437
2 -0.162980592469446
3 -0.22265702148705
4 -0.0950220080202581
5 0.0494193546125672
6 0.0265952491261698
7 0.0763662219747821
8 0.160929552717227
9 -0.018635131796938
10 -0.166773505486335
11 -0.0875321461681914
12 0.163524672212126
13 0.0853833116292758
14 -0.0992586730906817
15 -0.0845230631306517
16 -0.0119047647254736
17 -0.0746890698984212
18 0.137881339658262
19 -0.0309182497188642
20 -0.0731311933394466
};
\addplot [semithick, black, opacity=0.1]
table {%
0 1
1 0.00794717796398109
2 -0.285605554006185
3 -0.332444127866909
4 0.142488872108994
5 0.0857597888957774
6 0.0382538087786311
7 -0.0787116199946506
8 -0.0266944580165907
9 -0.0112620019841076
10 0.0303362660288215
11 -0.0494537413172426
12 0.0497578961292239
13 -0.0725137844202624
14 0.0469921630442169
15 0.019170703179332
16 0.0132178816855662
17 -0.0162913931351554
18 0.0329860893153507
19 -0.00962211596416861
20 -0.16053043159685
};
\addplot [semithick, black, opacity=0.1]
table {%
0 1
1 -0.0669088312636732
2 -0.121494681394725
3 -0.0506284357096252
4 -0.0904318728341608
5 0.123439203911846
6 -0.138712788220911
7 -0.0628405377160706
8 0.175824691094751
9 0.0803932484337716
10 -0.062763451497961
11 -0.0923101832885311
12 -0.0890049746925816
13 0.112402844354495
14 0.0577301803361284
15 0.0189382551389663
16 -0.106535142436192
17 -0.021098013091501
18 -0.0296159644192058
19 -0.0297000977285183
20 -0.0244922189052326
};
\addplot [semithick, black, opacity=0.1]
table {%
0 1
1 -0.0289334476412286
2 -0.0371894972434415
3 -0.0726308687344036
4 -0.0285554523904679
5 0.0626449623766436
6 0.0381859915770333
7 0.0682354465857766
8 -0.0611606826289788
9 -0.030054944270762
10 0.0301948956865804
11 0.0222311230099042
12 -0.0913058049845867
13 -0.0100525851127596
14 -0.0151580791012102
15 -0.0223619177157438
16 0.0599377172584969
17 -0.0438303003232502
18 -0.0255930280348295
19 -0.027260235837019
20 0.0962226596231007
};
\addplot [semithick, black, opacity=0.1]
table {%
0 1
1 -0.226311841812229
2 -0.180641701558987
3 0.4065643597294
4 -0.103469172427594
5 -0.0622214192400956
6 -0.189118651470458
7 0.100634353971269
8 0.1121814836416
9 -0.33211335030616
10 0.115847662289214
11 0.0536103530218683
12 -0.152419297148849
13 -0.0741697538649893
14 -0.0082815355538903
15 0.173640053906598
16 -0.175692832567371
17 -0.0229356471337429
18 0.246843877902668
19 -0.0807676128447487
20 -0.057936648530489
};
\addplot [semithick, black, opacity=0.1]
table {%
0 1
1 -0.372521786426247
2 -0.0384616391958357
3 0.0812620708613178
4 0.0549752267377332
5 -0.100733039771847
6 0.127830485053425
7 -0.18151260416569
8 0.0561099444052788
9 0.0341482406199791
10 0.0948240977872569
11 -0.138131551796359
12 0.0239889851363277
13 0.00689146656884494
14 0.0051157757372273
15 0.0374550609909453
16 -0.150379611210696
17 -0.00808675907074025
18 0.0874991600848617
19 -0.0407466283936408
20 0.0376303118987714
};
\addplot [semithick, black, opacity=0.1]
table {%
0 1
1 -0.00448701819908955
2 -0.231199993347321
3 -0.00124221342287727
4 0.237276175447468
5 -0.204305690288954
6 -0.173237627512099
7 -0.008944425479332
8 0.107967364951907
9 -0.0449387312812613
10 -0.0171065936927141
11 0.232006904267444
12 -0.0162576027867278
13 -0.159266513419525
14 -0.0584355916721877
15 0.195320629560583
16 -0.120795113516041
17 -0.28982322104643
18 0.0578320482308711
19 0.239980126618635
20 -0.0693421375742932
};
\addplot [semithick, black, opacity=0.1]
table {%
0 1
1 -0.0433684355047917
2 -0.0487928678180023
3 -0.00738949163512054
4 -0.0498417520375321
5 0.178200437275713
6 -0.191027056543833
7 -0.0248110784817403
8 0.0656535884591734
9 -0.0111533338695127
10 -0.0339588410637716
11 -0.0605183390185112
12 -0.0488109996280465
13 -0.0111404110604079
14 0.0213355449996405
15 0.0112300734516569
16 0.0920645241132397
17 0.0146521331497684
18 -0.0557448886761124
19 0.0652017463321048
20 -0.0439485530556988
};
\addplot [semithick, black, opacity=0.1]
table {%
0 1
1 -0.163455410116135
2 0.343044388663813
3 -0.0297718169855619
4 -0.048922621618455
5 -0.0991101353108829
6 0.0874531269580424
7 0.0440184365593921
8 0.187599355405072
9 0.0779957566316029
10 0.108862832013159
11 -0.0467244093230445
12 -0.00406579477192792
13 -0.133253508092291
14 0.0240349179802471
15 -0.122337018756798
16 0.0320938061684412
17 -0.0478556616073896
18 0.0221081645603134
19 -0.0109360874590551
20 -0.0351641928903406
};
\addplot [semithick, black, opacity=0.1]
table {%
0 1
1 -0.250758710607913
2 0.10965345501181
3 0.00566117503823132
4 -0.0812871120240028
5 -0.114516760596905
6 0.171702071062702
7 -0.123935635468788
8 -0.107201803637261
9 0.0770647361932264
10 -0.171335735416941
11 0.0509177161464788
12 0.0543371015289066
13 0.0180155959773447
14 -0.0771245596447215
15 0.239721207068286
16 -0.0801561504968692
17 -0.00362585952159994
18 0.0500439572196742
19 -0.137601142274313
20 -0.0502553480764143
};
\addplot [semithick, black, opacity=0.1]
table {%
0 1
1 -0.206896833893024
2 0.0824122464694541
3 -0.00417652873866134
4 -0.0507363044675079
5 0.0986925711063008
6 -0.115814488440578
7 0.137895528689792
8 -0.0341251570815424
9 -0.0181172206030184
10 0.0424345820618154
11 0.0828651252678441
12 -0.0523661072663565
13 0.00545118732429389
14 0.0876351178305135
15 -0.0328833117545594
16 -0.106504310979984
17 -0.0815746519664932
18 0.00292196329814088
19 0.0831012374044163
20 -0.0757175876589374
};
\addplot [semithick, black, opacity=0.1]
table {%
0 1
1 -0.243665459811793
2 -0.0493283657546035
3 0.252279494007066
4 0.0637436858796145
5 0.0254911831793902
6 -0.0165970016432746
7 0.0515598599282958
8 0.00844830741520678
9 -0.119439164707577
10 0.0381463598052343
11 -0.00353896604026797
12 -0.156798635239431
13 0.0707215090658362
14 -0.0185653381750384
15 0.0126260061765565
16 -0.0668908231315346
17 0.124386270908131
18 -0.0690818751655252
19 -0.0872969355399774
20 0.108010074960935
};
\addplot [semithick, black, opacity=0.1]
table {%
0 1
1 -0.167054232471641
2 0.122467784201223
3 0.00971390569608263
4 -0.0674646999643754
5 0.0150247827282347
6 -0.0186256302943036
7 -0.00864585193010529
8 0.00305330379770535
9 -0.046059688840698
10 0.0334373116824608
11 -0.0299121928179534
12 -0.00495688376720569
13 -0.012226431990398
14 -0.00936890712863715
15 -0.0147728511493568
16 -0.00925541913163663
17 -0.0186668926023538
18 -0.0119986820516031
19 -0.0144215568395463
20 -0.014522857427204
};
\addplot [semithick, black, opacity=0.1]
table {%
0 1
1 -0.0525252740596376
2 -0.0173156966047454
3 -0.113838484908201
4 0.0454802248997054
5 0.0408321731981845
6 0.026795070274056
7 -0.0483047552168884
8 -0.0268790416216285
9 -0.256415615029884
10 -0.100937742600627
11 0.0411384728903017
12 0.156688323513708
13 -0.00906052526935837
14 0.0159486712340033
15 -0.0175595970065352
16 0.0453712285254992
17 -0.0226228590652819
18 -0.0125740907840923
19 0.0684342883220515
20 -0.0552648915715344
};
\addplot [semithick, black, opacity=0.1]
table {%
0 1
1 0.0530969610506013
2 -0.157927867578236
3 -0.102500776190624
4 -0.0918268063190179
5 0.00112802724081328
6 0.0418589941399584
7 -0.0227386560594987
8 -0.00751171194581926
9 -0.00121099342401526
10 -0.0214448040694199
11 -0.000998453763533983
12 -0.00817019818342223
13 -0.0247927537264059
14 0.0501727237455035
15 -0.00778257124240635
16 -0.0369506858799693
17 -0.0256853467540028
18 -0.0159663384790877
19 -0.0148361854888354
20 0.00571984495027542
};
\addplot [semithick, black, opacity=0.1]
table {%
0 1
1 -0.197317336075113
2 -0.371834754443047
3 0.142468440651483
4 0.0487559461473432
5 0.0984110146686145
6 -0.183321190727531
7 0.00300142164449565
8 0.182176682504781
9 -0.129182042643539
10 0.0251344458725169
11 -0.1115018229563
12 0.00374075762863598
13 0.209298766860607
14 -0.0954933552885539
15 -0.0991182603771156
16 -0.031710134718031
17 -0.0834648039312657
18 0.0167684960756587
19 -0.0761642084173517
20 0.184003416345145
};
\addplot [semithick, black, opacity=0.1]
table {%
0 1
1 0.161144605064085
2 -0.0880244246470651
3 -0.132110026857255
4 -0.0611057527222309
5 -0.0718899006434076
6 -0.0149490599242375
7 0.239147084222876
8 -0.0392864078142003
9 0.0141011633207759
10 0.014136793222602
11 -0.0491178991063641
12 -0.135496670752027
13 0.0131728164408688
14 0.00345235675189517
15 -0.0364235028950441
16 0.0189743403829564
17 -0.0173277571672909
18 -0.0589022387370689
19 -0.0551705231381136
20 0.115257354162417
};
\addplot [semithick, black, opacity=0.1]
table {%
0 1
1 0.0979049785154578
2 0.15705293964912
3 0.125867669252105
4 0.0600298475663207
5 -0.0138921436493061
6 0.0495887176562624
7 0.214387022661582
8 -0.0753572277044115
9 0.156137403794394
10 -0.000697296413298019
11 0.0360352045408243
12 -0.00454629050159838
13 -0.0683772644111966
14 -0.0795395771001423
15 -0.051386757058227
16 -0.000620529745153694
17 -0.022448738723289
18 0.0326601795189014
19 0.00273066552674108
20 -0.0308614161633943
};
\addplot [semithick, black, opacity=0.1]
table {%
0 1
1 -0.146232767922019
2 0.00572371767000182
3 0.0243189334639758
4 -0.0708129896804771
5 0.110503021780403
6 -0.00865885862019069
7 -0.0394028385186762
8 -0.0137783010796362
9 0.107914437127805
10 -0.0333580128893897
11 0.0252804077543993
12 -0.0178924543595333
13 -0.0113715361418287
14 -0.0175849756446781
15 -0.019810767430963
16 -0.00580934983643276
17 -0.0232314583224257
18 -0.00723817081787167
19 -0.00276101757575968
20 -0.0240483205254508
};
\addplot [semithick, black, opacity=0.1]
table {%
0 1
1 -0.306804450188322
2 -0.0701036695224194
3 0.286375110791785
4 -0.196410305219629
5 -0.0612129905620811
6 0.0205617965182231
7 0.0799940947799854
8 -0.24490556077861
9 0.128773295509994
10 0.000664796103546809
11 -0.0972568651070081
12 -0.00274354581346627
13 -0.085615565518893
14 0.0969655955864523
15 -0.0478636427665981
16 -0.0827131540228659
17 0.358404856531003
18 -0.127463073892818
19 0.0290240100053377
20 0.251990007681315
};
\addplot [semithick, black, opacity=0.1]
table {%
0 1
1 -0.137181607219271
2 -0.0407958462777318
3 0.125074768456236
4 -0.215011132374768
5 0.00654380074640068
6 0.0215277322403175
7 -0.16142564206403
8 0.143240214076905
9 -0.0684735728801505
10 -0.137445629360545
11 0.111534705960444
12 -0.0294747097418663
13 -0.0783595967944004
14 0.171415801276018
15 -0.23398459327703
16 0.0223299006461711
17 -0.0630410918047106
18 0.101590379154715
19 0.0257796376278999
20 0.0652845250732789
};
\addplot [semithick, black, opacity=0.1]
table {%
0 1
1 -0.351870071204585
2 0.0428772655411139
3 -0.0695804548454552
4 0.0882799740177195
5 -0.0160785799845804
6 -0.260113135963844
7 0.231117566975614
8 -0.213101322237118
9 0.086986273049305
10 -0.0508659233381562
11 0.0413551042894673
12 -0.0949411159445335
13 0.0685705237788253
14 -0.021777376364054
15 0.0985059925007865
16 -0.0833252214604081
17 0.134351712548739
18 -0.0335408750716764
19 -0.0974020745250774
20 0.195755510707124
};
\addplot [semithick, black, opacity=0.1]
table {%
0 1
1 -0.0261066913023466
2 -0.185305488567702
3 0.0696763841245213
4 -0.33216415152946
5 -0.156614121896588
6 0.0203964481619933
7 0.110117621009581
};
\addplot [semithick, black, opacity=0.1]
table {%
0 1
1 -0.220485255187778
2 -0.0968795736425491
3 -0.0217124498444186
4 -0.0689520191965045
5 0.0154360238714588
6 0.170356633013246
7 -0.115012505441175
8 -0.0777871256683098
9 -0.0210040126995139
10 0.0312108668484377
11 -0.0223260919847338
12 0.145150677572014
13 -0.055752881002678
14 -0.173842064922609
15 0.0477413416318537
16 -0.0779440383140747
17 0.00684029281141231
18 0.161307746187172
19 0.0200169584356731
20 -0.226614114301985
};
\addplot [semithick, black, opacity=0.1]
table {%
0 1
1 0.137242259444259
2 0.00977516287719691
3 0.00208435435640186
4 -0.000383895343726048
5 0.0123390632484231
6 -0.00424513231981336
7 -0.00499312237507168
8 -0.00190459909849547
9 0.0125640216754835
10 -0.00220274991235776
11 0.0191041519777014
12 -0.00247900882050997
13 0.0111706151252287
14 -0.00757187820905915
15 -0.00241214125282797
16 -0.0055563968322148
17 0.000685196874890668
18 0.00567691114671591
19 -0.00457592071631874
20 0.0206091239606006
};
\addplot [semithick, black, opacity=0.1]
table {%
0 1
1 -0.336844319630527
2 0.0927425155248072
3 -0.126404853097833
4 0.14070800785981
5 -0.0617104688148258
6 -0.0961020057955904
7 -0.0556733779930819
8 0.0213410761847036
9 0.107459730449745
10 -0.193338486683025
11 0.199571177855078
12 -0.0854980559847617
13 0.173521266548425
14 -0.167183583206918
15 -0.0105458870091674
16 -0.00254424172375413
17 0.138771240825527
18 -0.0545579863940533
19 -0.0289893170229077
20 -0.0322787874190928
};
\addplot [semithick, black, opacity=0.1]
table {%
0 1
1 -0.359195899127815
2 0.089307238999858
3 -0.0173894802980565
4 0.0720776621368151
5 -0.0242321932421794
6 0.108271010571
7 0.184744584987339
8 -0.0467349071865546
9 0.0139601642909196
10 -0.081133260338634
11 0.221834455774926
12 -0.0824953341183217
13 0.12544816603452
14 -0.0851810101486056
15 0.0537286170134534
16 0.0206399889289321
17 -0.00247554195196216
18 0.0131942791137299
19 0.0203008576890222
20 0.024122329396716
};
\addplot [semithick, black, opacity=0.1]
table {%
0 1
1 -0.2779255266785
2 -0.00998186111746192
3 -0.0579735663103579
4 0.176709032613435
5 -0.215827556356222
6 0.136302513481536
7 -0.0163073759007946
8 0.0113117038103593
9 -0.0526217034424701
10 -0.0436749176213194
11 -0.0608856469465975
12 0.0745956717585095
13 -0.0342298995247819
14 0.0218824407336036
15 0.0142802679319561
16 -0.0539955221336942
17 -0.0250656583599301
18 0.122427946274069
19 -0.0482913232197703
20 -0.0715444804975198
};
\addplot [semithick, black, opacity=0.1]
table {%
0 1
1 -0.113822928094656
2 -0.00618281760286925
3 -0.110938634873975
4 -0.0835344309515052
5 0.0118582560577804
6 -0.116158094937574
7 0.174783930549028
8 -0.057629848831703
9 0.310570031431089
10 -0.0641375319795959
11 -0.102077830953619
12 -0.00852383305264338
13 -0.103629669609471
14 0.00636906325089163
15 0.0658131973636194
16 0.0546257901215428
17 -0.00370218040133641
18 -0.134032592300516
19 -0.0391340254447863
20 -0.0504142857130612
};
\addplot [semithick, black, opacity=0.1]
table {%
0 1
1 0.373030533512821
2 0.27780212632433
3 -0.0481714038620605
4 -0.0878575978612095
5 -0.0908648313882494
6 -0.0374818537916591
7 -0.0198493227339034
8 -0.0264439005205321
9 0.00202760772500388
10 -0.00133318119710101
11 0.0442291316358011
12 0.0472589415741099
13 -0.00112488264977743
14 -0.0312763284704878
15 -0.00766359057414728
16 -0.0732156784889811
17 -0.036377911845688
18 -0.0412192648903568
19 -0.0185192584995188
20 -0.00203303745803975
};
\addplot [semithick, black, opacity=0.1]
table {%
0 1
1 0.256517451063034
2 0.291530699109924
3 -0.0672939618656258
4 0.247384203655341
5 0.148764503880214
6 0.1860759671151
7 0.0104758993017293
8 0.00646390550323313
9 -0.0574861225460832
10 -0.0285376250993335
11 -0.0758288865551404
12 0.0423360263063708
13 -0.0482834723780367
14 -0.0363544177709517
15 -0.0318802536173998
16 -0.0419911086837756
17 -0.00563469696729071
18 -0.0997467246187839
19 -0.0633278833992247
20 -0.0879418108508342
};
\addplot [semithick, black, opacity=0.1]
table {%
0 1
1 0.19246558048276
2 -0.101042145676011
3 0.0765202309597355
4 0.0577630257161173
5 0.0351405410471785
6 0.0514918320265347
7 0.111801006955712
8 -0.00683876548999792
9 0.041021730493873
10 -0.0255467500169849
11 -0.0159345787338411
12 -0.00186132668463603
13 0.0158626293842179
14 0.0258852017792237
15 -0.0331363541492089
16 0.00664593314931354
17 0.0317221291876629
18 -0.110487978553398
19 -0.196972764371264
20 -0.0216944552321093
};
\addplot [semithick, black, opacity=0.1]
table {%
0 1
1 -0.223576614165879
2 -0.175705974304344
3 0.0931159467502395
4 0.0725730945049556
5 -0.025023384081677
6 -0.0535587516378812
7 0.102078233544086
8 -0.073562502504479
9 0.134528112732965
10 -0.0633335091398154
11 0.0147139937064465
12 0.15817334413564
13 -0.0782693174977582
14 -0.14454706576712
15 -0.170962892145149
16 0.183934749681996
17 0.0733021739265638
18 -0.106298258364653
19 -0.108867816744967
20 0.22373173701396
};
\addplot [semithick, black, opacity=0.1]
table {%
0 1
1 -0.171884370803442
2 0.146401174050719
3 0.0335740979852226
4 0.129256453298527
5 0.00861654964873683
6 -0.0816664685292347
7 -0.099456434626889
8 0.0287054579062964
9 -0.0228958547793797
10 -0.0275067325098615
11 -0.15724839345288
12 0.149765594173926
13 -0.00361280343806695
14 0.0162349753917182
15 -0.00745569648166144
16 -0.122216500994562
17 0.205260978844711
18 -0.150560793620628
19 -0.0507532625623924
20 0.0032870614919413
};
\addplot [semithick, black, opacity=0.1]
table {%
0 1
1 0.273281098581665
2 -0.184894550352062
3 0.0724807530105028
4 0.015181061621959
5 -0.070639674467809
6 -0.0767006354327088
7 0.041945367258019
8 0.128419303062529
9 -0.0418168588118411
10 -0.0856566354764711
11 -0.0250488324499308
12 -0.00678171396376105
13 -0.0260570927327723
14 -0.0498747343564548
15 -0.0714112565416431
16 -0.00137565697709669
17 -0.0127956068460588
18 -0.00090640401265921
19 0.0431460711844116
20 -0.0263973791584965
};
\addplot [semithick, black, opacity=0.1]
table {%
0 1
1 -0.226135869034557
2 -0.0551767899742362
3 0.0720065088060662
4 0.0109794427931478
5 -0.0871643490692179
6 -0.0207920165220586
7 -0.00128844862057122
8 -0.113813109832354
9 0.00914314241902683
10 -0.0701754748420765
11 -0.0280050534176779
12 -0.00388280016746651
13 0.0570059800226883
14 0.0433180187959834
15 -0.0338021994545211
16 -0.0213077338111637
17 -0.0324354922936214
18 -0.103006618539031
19 0.0175953107738817
20 0.118027370983248
};
\addplot [semithick, black, opacity=0.1]
table {%
0 1
1 0.179301452691429
2 -0.00134423962287232
3 -0.00748005984311495
4 -0.121729115234146
5 -0.18561397445578
6 -0.122705624552688
7 -0.00287867264128384
8 0.122021255567316
9 -0.0385083230314657
10 -0.0709172697232959
11 0.0648497686224909
12 -0.108651573671149
13 -0.0547852681642978
14 0.112024943623862
15 -0.0360633583147372
16 -0.0403508301968225
17 0.0713229739865727
18 -0.0239272005331813
19 -0.0754422161206708
20 -0.0432464116500019
};
\addplot [semithick, black, opacity=0.1]
table {%
0 1
1 -0.168659275113605
2 -0.0116063085568422
3 0.0997367961084389
4 -0.140367622368061
5 0.497822203658168
6 -0.047378879981994
7 -0.0513422227335957
8 0.0479838664552388
9 -0.0148386200920059
10 0.00343997769769807
11 -0.0152857629335674
12 0.00321428466468485
13 -0.0141240484249857
14 -0.0179358194673238
15 -0.0198280366165972
16 -0.0147973006996775
17 -0.00400555068006735
18 -0.0246003355607241
19 -0.0278804219260607
20 -0.0289276231437057
};
\addplot [semithick, black, opacity=0.1]
table {%
0 1
1 -0.255655799972436
2 -0.0101624233831929
3 -0.0054865770234237
4 -0.0262992700526626
5 0.132567516490763
6 -0.119768359897789
7 -0.0573047432120068
8 0.10315218077991
9 -0.0810782257139975
10 0.0205612015149671
11 -0.0458723298799552
12 0.0937064656618378
13 0.0400495326338424
14 0.0445739773276151
15 -0.0257344087243374
16 -0.0335648571491402
17 0.0297205905491276
18 -0.0344256321836737
19 0.0183466500377734
20 -0.0233631248640562
};
\addplot [semithick, black, opacity=0.1]
table {%
0 1
1 -0.106631474467916
2 0.145625307163738
3 0.128331932036192
4 0.0172285886537538
5 -0.013105828628694
6 0.01222011787099
7 0.00144521301097936
8 -0.0458431878214491
9 0.0994017737122848
10 -0.0598496645140453
11 0.00322700161216235
12 -0.00306654587068769
13 0.0394265189284615
14 -0.0669749879967736
15 0.00360297377448522
16 -0.0554302093472118
17 -0.00145972296207989
18 -0.00621313731860415
19 -0.00988868849190178
20 -0.0045281533534287
};
\addplot [semithick, black, opacity=0.1]
table {%
0 1
1 -0.117600649360518
2 -0.0677313175675657
3 -0.0447041557130772
4 0.0279382527429371
5 -0.0276639014723721
6 0.0177785630025729
7 -0.0019589969991801
8 -0.0194599973331841
9 -0.0745246895399793
10 0.160070027746607
11 0.068770525904882
12 -0.100383171891194
13 -0.167249062319771
14 0.0662916823053282
15 0.0397774371299429
16 -0.0322997307851702
17 -0.0207708835739664
18 0.0147386981909852
19 -0.0150001015686442
20 0.00626278683256236
};
\addplot [semithick, black, opacity=0.1]
table {%
0 1
1 -0.216906216844804
2 0.107925387713607
3 -0.0411477273059035
4 0.0445885057976457
5 -0.00654964849054859
6 -0.139062096152325
7 -0.0174113156137858
8 -0.0611012876169345
9 0.121736597781126
10 -0.18794542269333
11 0.103733345426614
12 -0.0897211305381172
13 0.0957479652837893
14 -0.054076239963407
15 -0.0208953195484191
16 0.198526921246623
17 0.0253276925230334
18 -0.0418270158865128
19 0.112368286709631
20 0.00318934745260936
};
\addplot [semithick, black, opacity=0.1]
table {%
0 1
1 -0.210486601788612
2 -0.274296004109124
3 0.106554147741786
4 0.0228270709660083
5 -0.0161118735112261
6 -0.0697158362557159
7 0.123474858097752
8 -0.247565074844452
9 0.0636251691937965
10 0.199924921804007
11 0.023867518457712
12 -0.00894145438155096
13 -0.0938923409154013
14 0.00609847725415133
15 -0.0378355151403324
16 0.0197274215161983
17 -0.0882128945197574
18 -0.0652162661827686
19 0.080568188280122
20 0.0419001476219803
};
\addplot [semithick, black, opacity=0.1]
table {%
0 1
1 -0.11666250180224
2 -0.222421726866959
3 0.030623919621223
4 0.0744185201307767
5 -0.108632261972883
6 -0.0998906477878193
7 0.440152489164631
8 -0.035321980101241
9 -0.0659818713405005
10 -0.00398857418280269
11 0.00306250671696818
12 -0.00362631800964694
13 -0.00578169084323374
14 -0.0107244836227613
15 -0.00528687493598141
16 -0.00473558631170854
17 -0.00590254933580063
18 -0.00411154789398491
19 -0.00712617755088891
20 -0.00694417568917718
};
\addplot [semithick, black, opacity=0.1]
table {%
0 1
1 -0.140187123307101
2 0.129656332347183
3 -0.0212850643708265
4 -0.121949882623094
5 0.409652163478656
6 -0.343582138187256
7 0.185514032297058
8 -0.184724110097596
9 -0.0801419685678716
10 0.144261730147604
11 -0.206716808885684
12 0.183632222558434
13 -0.276810039826841
14 0.028043734127762
15 0.0443248154343347
16 -0.0200902898959624
17 0.0655427455249476
18 -0.155971383583178
19 0.113700485596341
20 -0.0263582411907016
};
\addplot [semithick, black, opacity=0.1]
table {%
0 1
1 -0.312504158658497
2 -0.120544558548807
3 0.00268487614268155
4 -0.10321809116764
5 -0.0274662818047825
6 0.0483966822397896
7 -0.0897609716817288
8 0.247076073531227
9 -0.0145291100566372
10 -0.0819218089086046
11 -0.127641885332619
12 0.0230457921959503
13 0.155079834698945
14 -0.0617488553106426
15 -0.0501348201818191
16 0.0391421483005244
17 -0.00212940884902394
18 0.0789472470259278
19 -0.0802596325663544
20 -0.138275010348386
};
\addplot [semithick, black, opacity=0.1]
table {%
0 1
1 -0.365293189991042
2 -0.0358141711940571
3 -0.0127428092045799
4 -0.0530816678650676
5 0.130723354819713
6 -0.102781215319377
7 0.24042707491831
8 -0.18108019141179
9 0.342174995393799
10 -0.245874020047855
11 -0.125630615887583
12 0.316381491123872
13 -0.218784150683459
14 0.12570943787347
15 -0.0601852576847956
16 0.125241471238369
17 -0.030815763954549
18 -0.0958290612329513
19 -0.0113265149866248
20 -0.178981831122694
};
\addplot [semithick, black, opacity=0.1]
table {%
0 1
1 -0.0858316471693497
2 0.12656168412132
3 -0.180431451295151
4 0.00174212504968881
5 0.166600575699014
6 0.159244816970947
7 -0.081448097036775
8 -0.0637566680524556
9 -0.031937874841818
10 -0.0269875015219495
11 0.19205283039799
12 -0.011761726794953
13 0.0799764886269303
14 -0.0290969821860089
15 -0.0585215514634519
16 -0.0705172333593805
17 -0.0666568377413192
18 -0.0154846465642828
19 -0.000459512605838642
20 0.0225799857054194
};
\addplot [semithick, black, opacity=0.1]
table {%
0 1
1 -0.162616073521362
2 -0.172480623758719
3 0.192325479591996
4 -0.11096639596558
5 -0.055132718297856
6 0.0033418167569196
7 -0.0299452927050636
8 0.0902331247751869
9 -0.0433819706000169
10 -0.0649986953953874
11 -0.00308108453896382
12 0.153428741399934
13 -0.269622145015219
14 0.0562180165955537
15 0.271623129574623
16 -0.248224451118621
17 -0.0433523255585471
18 0.045852152204944
19 -0.145975868824382
20 0.0483306583755211
};
\addplot [semithick, black, opacity=0.1]
table {%
0 1
1 -0.365337831866265
2 0.12464353635525
3 -0.00239033922837094
4 0.0255802594330135
5 -0.0245327111202209
6 -0.0536024848908353
7 0.0203147552620951
8 0.0177793891499713
9 0.0476952257563674
10 -0.0483109788756027
11 0.0522441316161738
12 -0.130975244923463
13 0.0913158600385697
14 -0.134382360181337
15 0.0762450161991387
16 -0.0788900188000212
17 0.0964404130137029
18 -0.0817756005004461
19 0.170040398772036
20 -0.1215803057033
};
\addplot [semithick, black, opacity=0.1]
table {%
0 1
1 0.110919574397816
2 -0.00269495854804055
3 -0.0317183315592786
4 0.0129680957779208
5 0.141742717715578
6 -0.117866168011341
7 0.0405551809074311
8 0.163854407657761
9 0.0201409758557403
10 -0.0974057352487874
11 -0.0311853243264992
12 -0.126770068997374
13 0.0152158968590579
14 -0.043107855263641
15 0.0775693915408613
16 0.0395161941140533
17 -0.0624776323230539
18 -0.0567133292719552
19 0.00876762200766071
20 0.0198589172739594
};
\addplot [semithick, black, opacity=0.1]
table {%
0 1
1 -0.168463241243135
2 0.0440759749614944
3 -0.341718728193648
4 -0.072322950140769
5 -0.0612932095792343
6 -0.0819859063793646
7 0.202564111315369
8 0.122373531132498
9 0.0545718421944352
10 -0.240797012362759
11 -0.00241724524315392
12 -0.129076152928791
13 0.174488986467058
};
\addplot [semithick, black, opacity=0.1]
table {%
0 1
1 -0.0962865186256153
2 -0.0941362831710814
3 0.168972934188458
4 -0.0554223533378394
5 0.0138149944578775
6 -0.0456984433993193
7 -0.121435389349036
8 -0.035931821664842
9 0.00351950986038877
10 0.00217279458441406
11 -0.038211928496975
12 0.117022886190665
13 0.00602161695272987
14 -0.0343543934535797
15 -0.00870759833436759
16 -0.105002017115919
17 0.00211114056095273
18 0.0165740164424528
19 -0.0736252859690459
20 -0.0207384263435541
};
\addplot [semithick, black, opacity=0.1]
table {%
0 1
1 -0.448438181170531
2 0.0993618568903505
3 -0.302373357316509
4 0.323204960009777
5 -0.292018718253146
6 0.357983776272047
7 -0.310405935621412
8 0.208958175125278
9 -0.275584796736654
10 0.289396476903036
11 -0.222745954870033
12 0.273513570037796
13 -0.174436837466881
14 0.116726525875443
15 -0.257180531276326
16 0.165017883334078
17 -0.103263881579346
18 0.178476690426344
19 -0.171957330109857
20 0.18181759551559
};
\addplot [semithick, black, opacity=0.1]
table {%
0 1
1 -0.0911634048575201
2 -0.00251448531422308
3 0.0411825570388883
4 -0.0293707741675828
5 0.0186809236604194
6 -0.00708034592559476
7 -0.0397723885211904
8 -0.00821130085210239
9 -0.0254636903352772
10 -0.00658486034523578
11 -0.0173103027575081
12 -0.00330475660177552
13 -0.0240679441343455
14 0.0832517619740022
15 -0.0227109534649421
16 -0.0489941565704594
17 -0.00117362668078472
18 -0.00914469493907563
19 0.022005764638687
20 0.0493761645070342
};
\addplot [semithick, black, opacity=0.1]
table {%
0 1
1 -0.118027936797401
2 -0.128718331527738
3 0.0379394769885206
4 -0.11466896065095
5 -0.0322381550499939
6 -0.00968620264520483
7 -0.0321675028214561
8 -0.0456800077865215
9 -9.22728879304735e-05
10 0.112572250232294
11 0.0341662353831664
12 -0.0749868333075691
13 -0.0227999279515085
14 0.0138504032976518
15 -0.109691504419929
16 -0.0134791695047215
17 0.0461789165360701
18 -0.0583909683722963
19 -0.0158071652771136
20 0.0893645247325666
};
\addplot [semithick, black, opacity=0.1]
table {%
0 1
1 -0.334727209511689
2 -0.0546828079302288
3 0.134598928784058
4 -0.0768146139225116
5 0.24965838374561
6 -0.0390292929913422
7 0.0038834400124933
8 0.0242461813738347
9 -0.0426314856006652
10 0.155603453142291
11 -0.0847414618140739
12 0.0176962534825362
13 0.00731502793407988
14 0.0313981503309695
15 -0.0242777135872793
16 -0.043852382101926
17 0.00577072791054536
18 0.0668864641013769
19 -0.00439480696678287
20 -0.0516023394995264
};
\addplot [semithick, black, opacity=0.1]
table {%
0 1
1 -0.19095832452925
2 -0.0321515328986954
3 0.0132346785274178
4 0.10092051435558
5 0.0515386521977821
6 0.00710399879900194
7 -0.110706076425905
8 -0.131142548066072
9 0.0483851903171915
10 0.160956705140221
11 -0.18022940423307
12 -0.129622290230399
13 0.229460044109017
14 0.105550206197921
15 -0.124532345311507
16 -0.0297845093087268
17 -0.154425801666792
18 0.101513557396854
19 0.122564158884642
20 -0.193049409796769
};
\addplot [semithick, black, opacity=0.1]
table {%
0 1
1 -0.0559798770577577
2 -0.00636363285014842
3 0.00965815874665549
4 -0.0122992503671921
5 0.235736039652911
6 -0.00335059677143209
7 0.0134361222430652
8 0.0480846455160425
9 -0.0346684129695249
10 0.0172187689653797
11 0.00446196659973308
12 -0.0324229477121676
13 -0.0469578683731449
14 0.0149860785433758
15 -0.0333760329797851
16 0.0118837874892232
17 -0.0229941476525167
18 -0.0449650629666066
19 -0.0261423694115058
20 -0.00242686098641528
};
\addplot [semithick, black, opacity=0.1]
table {%
0 1
1 -0.326023811032745
2 0.0389599496122195
3 -0.186476058258328
4 -0.0326443083513257
5 0.0318654574005136
6 0.00933694155271934
7 0.00980528849840568
8 0.0215850556917716
9 -0.0648544067017387
10 -0.0118504472252042
11 0.024214327623691
12 -0.00937699163840662
13 -0.0311364083221413
14 0.0247027368690026
15 -0.0309415380946521
16 0.0592405353007543
17 -0.124109171070086
18 0.0767722615066153
19 0.0828214077131223
20 0.00594152284906325
};
\addplot [semithick, black, opacity=0.1]
table {%
0 1
1 0.130497396937292
2 -0.329315136696288
3 -0.292055326717189
4 -0.144398207326482
5 0.0632072708673281
6 0.102248024602083
7 0.00237611398987827
8 -0.0251731252734434
9 0.0141257273713411
10 0.00576096428678044
11 0.0821864451692076
12 -0.0106147444518873
13 -0.000266376325074109
14 -0.110853845622387
15 -0.189425499381103
16 0.0870297403589338
17 0.14637059736759
18 0.16872524110097
19 -0.0753577743933782
20 -0.115213096962541
};
\addplot [semithick, black, opacity=0.1]
table {%
0 1
1 -0.311110979607109
2 0.0500824740486132
3 -0.0580958245807643
4 -0.00196684431896509
5 -0.245114839724636
6 0.103532632107768
7 -0.0282304388924749
8 -0.0257193588531521
9 0.13277909065263
10 -0.0412335339526581
11 0.10600115211918
12 -0.0918791620938693
13 0.0672273402874042
14 -0.108764014036674
15 0.0164017653621037
16 -0.0456766866963306
17 0.0510514377749999
18 -0.0451837554789237
19 0.0533004008830966
20 -0.0559898682424433
};
\addplot [semithick, black, opacity=0.1]
table {%
0 1
1 -0.211195458061072
2 0.203649883089068
3 -0.0789897175835937
4 -0.0957440582360339
5 0.0889013749736669
6 -0.0297505772186215
7 0.0276511342078176
8 -0.063836009893926
9 0.0210083054574056
10 0.139724290383806
11 -0.0387223820592038
12 0.10645040342521
13 -0.0555026755455341
14 -0.112193555530895
15 -0.0542073253667705
16 -0.0393696286673931
17 -0.0391533180218947
18 -0.0153033293576693
19 -0.0316142541397485
20 0.0651177414503359
};
\addplot [semithick, black, opacity=0.1]
table {%
0 1
1 -0.137463046262668
2 0.224982547278326
3 -0.237544844426972
4 0.0908093945332873
5 -0.173682464641194
6 0.119958803203085
7 -0.0206945872809808
8 0.00224217390972852
9 -0.00118245937425843
10 -0.027384055378277
11 0.00997522291981171
12 -0.0572948774168705
13 -0.115348712140198
14 0.00810765737162219
15 -0.00397749350434435
16 0.0478898700646732
17 0.0117907965151272
18 -0.0710178661899021
19 -0.0578372160111676
20 -0.0482577400953274
};
\addplot [semithick, black, opacity=0.1]
table {%
0 1
1 -0.0361991026109397
2 0.0120199650516319
3 -0.00631860146911403
4 0.0164748102790744
5 0.0070583343660686
6 3.24672536724248e-05
7 -0.00228734484396505
8 -0.00411920800636802
9 -0.000635948372130221
10 -0.0218383270065515
11 0.00550319385089439
12 -0.0149234981686029
13 -0.000466023688190052
14 -0.0131759967263394
15 -0.0157357749858262
16 -0.00472503647517466
17 -0.0218910840299506
18 -0.00313899404287553
19 -0.0122729781033738
20 -0.0271100733693087
};
\addplot [semithick, black, opacity=0.1]
table {%
0 1
1 0.0395803843740636
2 -0.103695463997091
3 0.43826445634211
4 0.0328132005496487
5 -0.119080479368271
6 -0.0741265863805727
7 -0.101925939923945
8 -0.100313235016039
9 -0.270696896461745
10 -0.0900403961133513
11 0.0137446314541479
12 -0.164523675458954
};
\addplot [semithick, black, opacity=0.1]
table {%
0 1
1 -0.149201572057194
2 -0.129305674577171
3 -0.00942870936220359
4 0.195898664086114
5 -0.00346365808735333
6 -0.0739863856143124
7 -0.0498202099761481
8 0.220226085968249
9 0.0522318725916016
10 -0.0929155477193088
11 -0.0570559864224188
12 0.0244551671934056
13 -0.0610409224185089
14 0.0104957957906296
15 -0.0785410438892633
16 -0.074647640569539
17 0.0424591837731113
18 0.166000001666298
19 -0.0598894515629224
20 -0.0885387509383716
};
\end{axis}

\end{tikzpicture}

%% file: Graphs/acf/acf.200.tex
\begin{tikzpicture}

\begin{axis}[
height=\figH,
tick align=outside,
tick pos=left,
title={\(\displaystyle \Delta t=200\)},
unbounded coords=jump,
width=\figW,
x grid style={white!69.0196078431373!black},
xlabel={Lag},
xmin=-1, xmax=21,
xtick style={color=black},
y grid style={white!69.0196078431373!black},
ymin=-0.6, ymax=1,
ytick style={color=black}
]
\addplot [semithick, black, opacity=0.1]
table {%
0 1
1 0.370949788037107
2 -0.190056352029804
3 -0.350759914509754
4 -0.231053062931015
5 0.00812682778862193
6 -0.0427091185303734
7 0.057050890461707
8 -0.0116255593014016
9 -0.0773009632145167
10 0.0180918132000392
11 -0.0507143489706101
};
\addplot [semithick, black, opacity=0.1]
table {%
0 1
1 -0.0574478486051119
2 0.0341632243463766
3 -0.0436074673778659
4 0.0380235369522444
5 -0.0595148274423257
6 -0.0508135494447172
7 -0.0152627215238072
8 -0.0680279777293505
9 -0.0282319281668342
10 -0.027679063256237
11 -0.0831917347501464
12 -0.0626796493943701
13 -0.0757299936078547
};
\addplot [semithick, black, opacity=0.1]
table {%
0 1
1 0.192675238726782
2 -0.110972845007443
3 -0.0499627515524426
4 0.115504136608697
5 -0.171976619696183
6 -0.250520866380486
7 -0.0173836326772881
8 -0.0219593556637366
9 -0.0696376524550806
10 -0.0954204014878783
11 -0.02034525041494
};
\addplot [semithick, black, opacity=0.1]
table {%
0 1
1 -0.321597092895647
2 -0.101983916536523
3 -0.0133957933031388
4 -0.11646809649468
5 -0.0184492397264988
6 0.262064597081489
7 -0.207449730383925
8 0.0172792722589231
};
\addplot [semithick, black, opacity=0.1]
table {%
0 1
1 -0.545452036093361
2 0.559181420605196
3 -0.534511707583322
4 0.32627560503247
5 -0.456761883825206
6 0.344323556055599
7 -0.330476868156561
8 0.287717020103856
9 -0.279029763114986
10 0.363158345198818
11 -0.2574572745476
12 0.202527296445841
13 -0.129492821142948
14 0.0894348604727653
15 -0.0474066991452343
16 0.0327769373209936
17 -0.0352602531109068
18 -0.0228826413913246
19 -0.0242955750513342
20 -0.00349106256223668
};
\addplot [semithick, black, opacity=0.1]
table {%
0 1
1 0.372468189853055
2 0.0916456899079215
3 0.130569987239563
4 -0.106242955904892
5 -0.0323791278604797
6 -0.0443037700278029
7 -0.176135926369585
8 -0.130650059841467
9 -0.12261722589013
10 -0.297314718747251
11 -0.190259987545635
12 0.00521990518670433
};
\addplot [semithick, black, opacity=0.1]
table {%
0 1
1 -0.607496061381779
2 0.150045054610893
3 0.299588345532992
4 -0.435118840716181
5 0.235385456940289
6 -0.142403954986214
};
\addplot [semithick, black, opacity=0.1]
table {%
0 1
1 0.126410434551961
2 -0.126422506033598
3 0.0398511162770423
4 -0.0877227062179809
5 -0.1612530954125
6 -0.254852127627406
7 -0.0360111155375198
};
\addplot [semithick, black, opacity=0.1]
table {%
0 1
1 -0.0199699355791302
2 0.0408336028402928
3 -0.174875261145732
4 -0.515364561373024
5 -0.0117811981456072
6 -0.0603001492427165
7 -0.0167944392187365
8 0.200588308871348
9 0.0925545827984318
10 0.00753605583850971
11 0.104796101349788
12 -0.147223106993424
};
\addplot [semithick, black, opacity=0.1]
table {%
0 1
1 -0.167906399464833
2 -0.1145715720027
3 -0.054090196538451
4 -0.209596512004029
5 -0.000458976298483706
6 -0.0984903422391924
7 -0.0691998630703412
8 0.102009406093647
9 0.297838273263955
10 -0.132982545451552
11 -0.00257058638599634
12 -0.0274826259831312
13 -0.0224980599188917
};
\addplot [semithick, black, opacity=0.1]
table {%
0 1
1 -0.11472799458887
2 -0.287803213171613
3 -0.165957278801305
4 0.216175225058214
5 -0.0218605683405557
6 0.0332561827718398
7 -0.320281658961127
8 0.0931205180850885
9 0.104513744819334
10 -0.0867779152744802
11 -0.0303901399330868
12 0.0759613383417377
13 0.00477175999482446
};
\addplot [semithick, black, opacity=0.1]
table {%
0 1
1 -0.30398945005116
2 0.171729502400564
3 -0.135011647273576
4 0.136176863571005
5 -0.192489251075566
6 0.253898490491851
7 -0.00574422784643809
8 -0.117637152461183
9 0.0185489789499427
10 0.000816768300372933
11 -0.072014799145035
12 0.0989170979840197
13 0.00932430195665979
14 -0.0602391824324502
15 -0.06587070246047
16 0.0787461166242675
17 -0.118746290555696
18 -0.0235557667706043
19 -0.0285216324476345
20 0.0122937419513593
};
\addplot [semithick, black, opacity=0.1]
table {%
0 1
1 -0.318749997421986
2 -0.27806306025066
3 0.134296790893528
4 -0.196706047182693
5 0.127618225080041
6 0.0893540394733543
7 0.0924597846232258
8 -0.134014965508355
9 -0.0920394324257719
10 0.0661186024309439
11 0.0097260602883716
};
\addplot [semithick, black, opacity=0.1]
table {%
0 1
1 0.129536204619323
2 -0.18325881882285
3 -0.224405700421364
4 -0.0602171336195978
5 0.192119326916502
6 0.0538141799398089
7 0.00077524464916774
8 -0.156202856317022
9 -0.1379476026496
10 -0.114212844294367
};
\addplot [semithick, black, opacity=0.1]
table {%
0 1
1 0.187761429713413
2 -0.289907407728755
3 -0.238991254366653
4 0.0725170295778389
5 0.271552035845694
6 0.0422743882748323
7 -0.209477972735124
8 -0.104775766518678
9 0.0971357428900748
10 0.012482931682716
11 -0.117669728286504
12 -0.174029392997969
13 -0.100602378783442
14 0.110930822302804
15 0.0969671477588064
16 -0.156167626629055
};
\addplot [semithick, black, opacity=0.1]
table {%
0 1
1 -0.13187963437496
2 0.07295370834885
3 -0.0285275733437305
4 0.173154348765196
5 -0.126580780102736
6 -0.0908155226538257
7 -0.060592244273973
8 -0.230493812719505
9 -0.0772184896453158
};
\addplot [semithick, black, opacity=0.1]
table {%
0 1
1 -0.21424292276398
2 -0.069454337659181
3 -0.245145394273193
4 -0.12979327383946
5 0.404857708853391
6 -0.175983222684975
7 0.0600384854473797
8 -0.105127821841283
9 -0.0143109480770606
10 0.0174049961850214
11 -0.0282432693466587
};
\addplot [semithick, black, opacity=0.1]
table {%
0 1
1 -0.12435433855834
2 -0.00158740644226306
3 -0.271768114809739
4 -0.330859656497177
5 0.243417623929648
6 0.038751981782691
7 0.269042816083953
8 -0.0198534518321217
9 -0.176088626315284
10 -0.0893595765937824
11 -0.0373412507475856
};
\addplot [semithick, black, opacity=0.1]
table {%
0 nan
};
\addplot [semithick, black, opacity=0.1]
table {%
0 1
1 0.0427643705540356
2 -0.020079304945309
3 0.229321859350511
4 -0.111903418229816
5 -0.151902731336162
6 -0.245864304400446
7 -0.0107108958760512
8 0.00652431888491419
9 -0.0330008732487778
10 0.107411932916012
11 0.0170833942539006
12 -0.0453644163454673
13 0.0035500733924807
14 -0.0691056515691453
15 -0.0745333139767723
16 -0.0159768794082702
17 -0.0322216042632324
18 -0.0373648070505114
19 -0.0253978187701855
20 -0.023583147763843
};
\addplot [semithick, black, opacity=0.1]
table {%
0 1
1 -0.0312140121878432
2 -0.0217836482020864
3 -0.0552382156398595
4 -0.0124676403361029
5 0.0127498146412819
6 -0.0377416833179996
7 -0.020042640731195
8 -0.0366151943012871
9 -0.0413651743499965
10 -0.0333757975575507
11 -0.0387363602766215
12 -0.03361849707479
13 -0.0438396421737708
14 -0.0530923408535125
15 -0.0536189676386661
};
\addplot [semithick, black, opacity=0.1]
table {%
0 1
1 -0.261279963094386
2 0.274589216120483
3 -0.039193784742072
4 -0.388678674054401
5 0.290492168769992
6 -0.0500108766742657
7 0.0529554773798523
8 0.180186019287017
9 -0.37048587671484
10 0.0488867062863964
11 -0.237460412563776
};
\addplot [semithick, black, opacity=0.1]
table {%
0 1
1 -0.0433960846836822
2 -0.010160046377795
3 -0.072845903060504
4 -0.0819440661051994
5 0.0253208332665033
6 -0.13936859885385
7 0.00757739835989401
8 -0.167097037151141
9 -0.0180864953942258
};
\addplot [semithick, black, opacity=0.1]
table {%
0 1
1 -0.204021771073788
2 0.0355177538060841
3 -0.147616834716208
4 0.0228129662606762
5 -0.00552615660591019
6 -0.0311586169152845
7 -0.0325181678003341
8 -0.0053261662219047
9 -0.0101756418356153
10 -0.0538675488730698
11 0.000584456753887876
12 -0.0617123143354634
13 -0.0224112023511996
14 0.00658524623956531
15 0.0064944813525356
16 0.0023395163160286
};
\addplot [semithick, black, opacity=0.1]
table {%
0 1
1 -0.0308737974934272
2 0.00575482635462617
3 0.0128297886326955
4 -0.0161478232226979
5 -0.00610294130479146
6 -0.109075367939469
7 -0.114401548757698
8 -0.0602029852644887
9 -0.10046018222497
10 -0.0813199687797801
};
\addplot [semithick, black, opacity=0.1]
table {%
0 1
1 0.190818574818166
2 0.0336793348302157
3 -0.10892955089309
4 -0.0281463881079956
5 0.0334170565986593
6 -0.0342085140463365
7 -0.095047763652891
8 -0.163037031408281
9 -0.0814352775662814
10 -0.097139395101384
11 -0.149971045470782
};
\addplot [semithick, black, opacity=0.1]
table {%
0 1
1 0.0500567246504902
2 -0.059897378296763
3 -0.0524770070010125
4 -0.0506881129570725
5 -0.0640812123770675
6 -0.0532886244919312
7 0.00714743942046079
8 -0.111868975317662
9 -0.0981702662493705
10 -0.0667325873800717
};
\addplot [semithick, black, opacity=0.1]
table {%
0 1
1 -0.374044280286289
2 0.0251753655483005
3 0.0436324028236136
4 -0.328722310223223
5 0.112325090598702
6 -0.0925323088157378
7 -0.059722601931851
8 0.384798006478886
9 -0.249771233748591
10 0.0354627613631712
11 0.00339910819301883
};
\addplot [semithick, black, opacity=0.1]
table {%
0 1
1 -0.25497915154253
2 -0.418084961522994
3 0.280997552965818
4 -0.111604902599043
5 -0.114779687031974
6 0.0434917009822425
7 0.11294882230781
8 0.11060830728879
9 -0.0428422157015893
10 -0.17115007144848
11 0.0208807617234558
12 0.0473275407656846
13 -0.00281369618719042
};
\addplot [semithick, black, opacity=0.1]
table {%
0 1
1 -0.125286572010415
2 0.0603073582235642
3 0.116772567212467
4 -0.148584953495747
5 -0.0655420530415449
6 -0.0337550769400554
7 -0.10530699865305
8 -0.198743215327835
9 0.0224041864167126
10 -0.0642585500371418
11 -0.0830528663140963
12 0.349002462316611
13 -0.223956288349469
};
\addplot [semithick, black, opacity=0.1]
table {%
0 1
1 0.0173077732069229
2 -0.170737378037641
3 0.000100989089870159
4 0.0410969322488315
5 -0.150705812789234
6 -0.0650341914119173
7 -0.0776284990054671
8 -0.0943998133013649
};
\addplot [semithick, black, opacity=0.1]
table {%
0 1
1 0.183663666967182
2 0.132477959333992
3 -0.0517804437668805
4 -0.375511772527303
5 -0.170814920997256
6 -0.0917157447675758
7 0.0471625869056507
8 0.0467875155388699
9 0.0128770315649316
10 -0.0711301971104934
11 0.0506187773321075
12 -0.300050010164998
13 -0.00457554349673194
14 0.0225747056384395
15 -0.0691400334280083
16 0.138556422978073
};
\addplot [semithick, black, opacity=0.1]
table {%
0 1
1 -0.275068587836742
2 -0.0126897999299194
3 -0.177877061394687
4 0.0829409555061679
5 0.072628460367716
6 -0.0497294334357645
7 0.0270698091774624
8 -0.0704011177853734
9 0.0394886136572775
10 -0.14240493045321
11 -0.0937074780925046
12 0.0959196747264743
13 0.0321228404546354
14 -0.0134478384113038
15 -0.0359600682537338
16 0.0247897295369903
17 -0.00358698603884851
18 -8.67817946359647e-05
};
\addplot [semithick, black, opacity=0.1]
table {%
0 1
1 -0.611662225510041
2 0.00681936588388304
3 0.203161803106546
4 -0.189392504656996
5 0.172568685823599
6 -0.109744443968216
7 0.0282493193212244
};
\addplot [semithick, black, opacity=0.1]
table {%
0 1
1 -0.32422299953701
2 0.149906056747129
3 -0.342181496063879
4 0.0497813558883963
5 -0.102784961912961
6 0.317703368863565
7 -0.130310566687244
8 -0.0394186266980714
9 -0.129759395433301
10 0.125371202189966
11 -0.0412935638877604
12 0.197938259841652
13 -0.229808826848896
14 -0.0714910468898376
15 -0.0212269356313736
16 0.11942256866721
17 0.0804315963642996
18 0.130273029935774
19 -0.151730244522442
20 -0.0865987743852166
};
\addplot [semithick, black, opacity=0.1]
table {%
0 1
1 -0.3914371136467
2 -0.047959102803789
3 0.233726130024677
4 -0.316605648344512
5 0.145502139995433
6 0.0495367289913317
7 -0.146388579996778
8 -0.117643829651985
9 0.0912692754323213
};
\addplot [semithick, black, opacity=0.1]
table {%
0 1
1 0.0165089385040819
2 -0.000412223418712047
3 -0.0968622387359209
4 -0.0691945958045558
5 0.151391324233557
6 -0.0967554421852474
7 -0.0220975086264293
8 -0.02436632956507
9 -0.100804942513752
10 -0.0146622496629872
11 0.0172489876506659
12 -0.116349263173391
13 -0.0619943859259887
14 -0.0898307009185115
15 0.00818063014226107
};
\addplot [semithick, black, opacity=0.1]
table {%
0 1
1 -0.176505806528814
2 -0.513220075821688
3 0.280258530691181
4 0.123260256402024
5 -0.0353483803290484
6 -0.16160736454243
7 -0.079576655445022
8 0.0627394955737979
};
\addplot [semithick, black, opacity=0.1]
table {%
0 1
1 -0.190563654741312
2 0.12400071131172
3 0.271650680134844
4 -0.336584155582231
5 0.277492940477893
6 -0.0939062091711666
7 -0.113728225850713
8 -0.0540413405265317
9 -0.181764347355479
10 -0.033846424413867
11 -0.168709974283158
};
\addplot [semithick, black, opacity=0.1]
table {%
0 1
1 0.0966407397979675
2 -0.20693090588331
3 0.0739340472441398
4 -0.244207207325002
5 -0.0775080934251424
6 0.0247309431527834
7 -0.150296679554585
8 -0.0163628440068519
};
\addplot [semithick, black, opacity=0.1]
table {%
0 1
1 -0.227501817894361
2 -0.149760396541
3 -0.0157908992313267
4 0.203509040660222
5 -0.265215011227563
6 -0.276063637702708
7 0.230822721936737
};
\addplot [semithick, black, opacity=0.1]
table {%
0 1
1 -0.277464389551141
2 0.487686954683045
3 -0.157772016041873
4 -0.0712946830421401
5 -0.140832321569875
6 -0.148963736219765
7 -0.129923064115176
8 -0.0614367441430749
};
\addplot [semithick, black, opacity=0.1]
table {%
0 1
1 -0.137404337015094
2 -0.456196569631749
3 0.15859646884809
4 0.202459060524275
5 -0.335329430996479
6 -0.103082306959862
7 0.378273915633296
8 -0.04313818294302
9 -0.0918390680715478
10 -0.0723395493879093
};
\addplot [semithick, black, opacity=0.1]
table {%
0 1
1 -0.302339498058218
2 -0.118011253489565
3 0.336575746279204
4 -0.176279055252819
5 -0.0536484741798674
6 -0.0928572742235413
7 0.0599809376572867
8 -0.0802099425109538
9 -0.109435348006736
10 0.329238842802516
11 -0.165457278767476
12 -0.0507447499380565
13 0.161117483620629
14 -0.0903814523933219
15 -0.0475722759230802
16 -0.0263522062752892
17 -0.0270015342330745
18 -0.0153435690349785
19 -0.0350932849423041
20 0.00381418686964712
};
\addplot [semithick, black, opacity=0.1]
table {%
0 1
1 0.270612859720553
2 -0.217021765785195
3 -0.376610165880732
4 0.0220182198288683
5 0.264564852775554
6 -0.00734239149147603
7 -0.223825236469512
8 -0.232396372698062
};
\addplot [semithick, black, opacity=0.1]
table {%
0 1
1 -0.114933241060067
2 -0.0667932257614028
3 -0.0457546483458583
4 -0.0679524181589503
5 -0.0609229825619738
6 -0.00927823016077649
7 -0.0626029636624679
8 -0.0717622902885029
};
\addplot [semithick, black, opacity=0.1]
table {%
0 1
1 -0.0140853210871952
2 -0.0772678576957612
3 -0.279803677624149
4 0.0420242295975019
5 -0.052816563257643
6 0.0734355992827277
7 -0.217673343932396
8 -0.0467956451223864
9 -0.0199650669962583
10 0.072150262008225
11 0.0207973848273344
};
\addplot [semithick, black, opacity=0.1]
table {%
0 1
1 -0.303925068557441
2 -0.11432198183082
3 0.101857417937774
4 -0.156072392486906
5 0.0924613888690516
6 -0.134604631954247
7 -0.224536430798527
8 0.239141698821115
};
\addplot [semithick, black, opacity=0.1]
table {%
0 1
1 -0.0148880271844631
2 -0.392729922529673
3 -0.133361085175113
4 -0.129058292164498
5 0.393172962325539
6 0.000192304921189817
7 -0.254910406723984
8 0.0315824665310021
};
\addplot [semithick, black, opacity=0.1]
table {%
0 1
1 -0.122105923874653
2 0.0744287481425508
3 -0.108690237306801
4 -0.0808401796445794
5 0.0248526116628201
6 0.0132705875420749
7 -0.167457954639628
8 0.169331054987132
9 -0.225216829592973
10 -0.0775718772759434
};
\addplot [semithick, black, opacity=0.1]
table {%
0 1
1 0.0537133717376899
2 -0.0141158360693677
3 -0.0440740041405433
4 -0.0164242590169036
5 -0.0484618299313308
6 -0.0400095130383795
7 -0.0210852517390584
8 -0.0542574518498378
9 -0.073888667476194
10 -0.0753751572361477
11 -0.0807323794832658
12 -0.0852890217566613
};
\addplot [semithick, black, opacity=0.1]
table {%
0 1
1 -0.386147879651521
2 -0.0853263612688995
3 0.386345458107809
4 -0.0370762324712444
5 -0.298646088691401
6 0.262892450287114
7 -0.120280564408116
8 -0.0131254705777522
9 -0.142295956162358
10 0.0959288254293627
11 -0.127256909613674
12 -0.0350112709793187
};
\addplot [semithick, black, opacity=0.1]
table {%
0 1
1 -0.157760723607863
2 -0.0545232795542359
3 0.226153738129457
4 -0.273942807214141
5 0.00396952621631914
6 -0.239057959790613
7 -0.0232513924617065
8 0.0834870808676084
9 -0.061417742194548
10 -0.0159277985289123
11 0.0122713581386355
};
\addplot [semithick, black, opacity=0.1]
table {%
0 1
1 -0.234259721551853
2 -0.0193358585612897
3 0.0718428870562225
4 -0.2341416375761
5 0.311626057450627
6 -0.141170680397878
7 -0.0841451723054072
8 -0.0689741218315014
9 -0.10144175228282
};
\addplot [semithick, black, opacity=0.1]
table {%
0 1
1 -0.452173034485922
2 -0.0999683747968661
3 0.0340650284169511
4 0.0496290591659835
5 -0.0598390717367849
6 0.0177911325211
7 0.0104952609155381
};
\addplot [semithick, black, opacity=0.1]
table {%
0 1
1 -0.5
};
\addplot [semithick, black, opacity=0.1]
table {%
0 1
1 -0.183523216110936
2 -0.13075802792109
3 0.00548671379121189
4 0.15254162290917
5 -0.586671991938061
6 -0.0602228156480145
7 0.247731057823023
8 0.0223229793280254
9 -0.0670471708014136
10 0.168921642118607
11 0.0810935041164227
12 -0.149874297666945
};
\addplot [semithick, black, opacity=0.1]
table {%
0 1
1 -0.00145853825296462
2 0.00422714350621957
3 -0.00696396168789163
4 -0.0112650255373802
5 -0.0096100693596692
6 -0.0313231962141612
7 -0.0357857138056523
8 -0.0124284995412473
9 -0.0471218764411931
10 -0.0414708132072371
11 -0.0474427714271599
12 -0.0410013430321276
13 -0.0432672181999884
14 -0.0616433198523386
15 -0.0511308535432167
16 -0.0623139434039918
};
\addplot [semithick, black, opacity=0.1]
table {%
0 1
1 -0.189966881923697
2 -0.348843062639986
3 0.300126299322644
4 0.112457512028263
5 -0.272874742725473
6 0.112544561065758
7 -0.0317781480967825
8 -0.08583430440582
9 -0.0186920590875195
10 -0.0714487785095757
11 -0.0454679584002747
12 0.192595658119616
13 -0.199927163459611
14 -0.116371722681841
15 0.163065542172652
16 0.0610176230138948
17 -0.0704056552652259
18 0.00963888489961596
19 0.000164396573362051
};
\addplot [semithick, black, opacity=0.1]
table {%
0 1
1 0.381072782433197
2 0.242530743103128
3 -0.157659776517271
4 -0.105222027694519
5 -0.0234862634081112
6 -0.122845925566905
7 -0.0430808209931351
8 -0.195992437954561
9 -0.242399483676198
10 -0.185025150578086
11 -0.114496411862376
12 0.00758125784308916
13 0.0109504587665792
14 0.0200618286192358
15 0.0280112274859317
};
\addplot [semithick, black, opacity=0.1]
table {%
0 1
1 0.0543416577782177
2 -0.000382714305501538
3 0.170611730804048
4 -0.178600861556469
5 0.153692222817136
6 -0.264625710906606
7 -0.195003733007302
8 0.0333166910746817
9 -0.221849609319547
10 -0.143721910665282
11 -0.138498821849315
12 -0.0358480270374969
13 0.0890527990648135
14 0.057633600387771
15 0.0781163683258618
16 0.0417663183949902
};
\addplot [semithick, black, opacity=0.1]
table {%
0 1
1 -0.390171677073904
2 0.200751539652033
3 -0.17575942172935
4 -0.0979792725011983
5 -0.0968159275095745
6 -0.0624002595940844
7 0.0916965988160619
8 0.118716087184916
9 -0.0525364187455429
10 0.270261627049731
11 -0.27289342057721
12 0.0266609605491167
13 -0.0149055342575547
14 -0.0145602590098226
15 -0.00804263426993866
16 -0.0711311892118975
17 0.0491092012282182
};
\addplot [semithick, black, opacity=0.1]
table {%
0 1
1 -0.022279677053004
2 -0.0174677520585782
3 -0.0149413739671381
4 -0.0446826729302401
5 -0.0448083736392984
6 -0.0540879671163509
7 -0.0560710508504673
8 -0.0776254407746538
9 -0.0817768460676793
10 -0.0862588455425898
};
\addplot [semithick, black, opacity=0.1]
table {%
0 1
1 0.283268483722201
2 0.110134169643044
3 -0.0579476335745344
4 -0.072792878154429
5 -0.108994534550919
6 -0.0656760315423822
7 -0.0768693046284355
8 -0.106880951892206
9 0.0440651344604693
10 -0.0514843386842048
11 -0.0965370622070207
12 -0.0522332097053303
13 -0.118544718919023
14 -0.129507123967229
};
\addplot [semithick, black, opacity=0.1]
table {%
0 1
1 0.296802121854912
2 0.000296662638948891
3 -0.0712126245055088
4 -0.159748228904525
5 -0.0164326904960334
6 -0.0679002624720012
7 -0.228963414146544
8 -0.252841563969248
};
\addplot [semithick, black, opacity=0.1]
table {%
0 1
1 0.553623166029228
2 0.286308407166942
3 0.0998786373503742
4 -0.19384267194495
5 -0.360345729451781
6 -0.389384799123715
7 -0.246980692657978
8 -0.12118429472221
9 -0.128072022645909
};
\addplot [semithick, black, opacity=0.1]
table {%
0 1
1 0.137563143513916
2 -0.113537510483844
3 0.0273132265068213
4 -0.184081682340448
5 -0.0986700783588345
6 -0.187353058804064
7 -0.137432887758607
8 -0.0425657602955863
9 -0.00206888112190139
10 0.25773770072651
11 0.159471970169344
12 0.101657799218444
13 0.0630308997060489
14 -0.241262734402732
15 7.77280037361637e-05
16 -0.0439003544840882
17 -0.151981500220238
18 -0.0439980195744757
};
\addplot [semithick, black, opacity=0.1]
table {%
0 1
1 0.244691859310937
2 -0.102401613596308
3 -0.296241493381911
4 -0.184326209822693
5 0.0444508893530996
6 0.100869266141379
7 -0.0893414168167371
8 -0.217701281187766
};
\addplot [semithick, black, opacity=0.1]
table {%
0 1
1 -0.325046546336138
2 -0.192068312787511
3 0.219345665507086
4 -0.255959605176717
5 0.0510310240696051
6 0.202549828884968
7 -0.136976946914814
8 -0.0368199069993567
9 -0.0193066661785325
10 0.0415062126007474
11 -0.1774570711681
12 0.0516864071634342
13 0.150661261496919
14 -0.0525702980666864
15 -0.0332978088618665
16 -0.0132982811435473
17 -0.049660933427045
18 0.0780581570832854
19 0.0396135003981544
20 -0.0419896801438848
};
\addplot [semithick, black, opacity=0.1]
table {%
0 1
1 -0.0934301326006843
2 -0.0403001241870572
3 -0.0247858129847702
4 -0.00603762731242985
5 -0.0378728202691244
6 -0.0397092274076632
7 -0.0402738891990528
8 0.00337821594879246
9 -0.0588695540688904
10 -0.0318609351935626
11 -0.0299254867381029
12 -0.0525689053880559
13 -0.0477437005993988
};
\addplot [semithick, black, opacity=0.1]
table {%
0 1
1 0.213033226475621
2 -0.0317254449572183
3 -0.0327612920149965
4 -0.0635670988167615
5 -0.0857016852948792
6 -0.144859348543587
7 -0.257339999652853
8 -0.0886447361374793
9 -0.00843362105784664
};
\addplot [semithick, black, opacity=0.1]
table {%
0 1
1 0.214023042913
2 -0.0363803156992529
3 -0.154590453703939
4 -0.324578428365266
5 -0.0246339834750127
6 -0.157551636745643
7 0.00390510029164197
8 0.0242320740549341
9 -0.0444253992704628
};
\addplot [semithick, black, opacity=0.1]
table {%
0 1
1 0.001180518154501
2 -0.0883736364693245
3 -0.0407977155651517
4 -0.126371263263005
5 -0.0220058200831921
6 0.124612387129141
7 -0.026730884289358
8 0.0365597387271494
9 0.0961199814513059
10 -0.112538644756184
11 -0.00167869062871811
12 -0.0330703821406667
13 0.0189727120523106
14 -0.0586367013847506
15 -0.041310097670228
16 0.0575663709775881
17 -0.00151980016939036
18 -0.0138955097935277
19 -0.0217699374494586
20 -0.060618038041655
};
\addplot [semithick, black, opacity=0.1]
table {%
0 1
1 -0.275355023053941
2 -0.0952453671198719
3 0.25205145087214
4 -0.119099409839169
5 -0.330954057950932
6 0.221289463280707
7 -0.197591968354781
8 0.0449049121658476
};
\addplot [semithick, black, opacity=0.1]
table {%
0 1
1 -0.14045421344168
2 -0.0469370460261894
3 0.0390415007928202
4 0.035827260836819
5 0.0811719921588495
6 -0.044572316219644
7 0.11080555432164
8 -0.153284222010254
9 0.050756195517357
10 -0.111859663485945
11 0.0234169598025813
12 -0.0760841920127461
13 0.00710597392863218
14 -0.0733805586515388
15 0.0004998539005781
16 0.22144560366632
17 -0.106225603185114
18 -0.0298208035203944
19 -0.0102815853452531
20 -0.0455224657494496
};
\addplot [semithick, black, opacity=0.1]
table {%
0 1
1 -0.0314660526821877
2 -0.330810974454062
3 0.00597043368206485
4 -0.0903493942381215
5 -0.12068116093588
6 0.0380490331811377
7 0.175461629096295
8 0.113738046881034
9 -0.255595602863015
10 -0.0325304585201031
11 0.176714505200315
12 -0.148500004347477
};
\addplot [semithick, black, opacity=0.1]
table {%
0 1
1 -0.350207107512889
2 -0.104652949188162
3 0.0356487849002816
4 -0.0286945339991121
5 -0.011019442221637
6 0.00324012718176296
7 -0.0077827571634064
8 -0.0149682153197411
9 -0.0215639066770966
};
\addplot [semithick, black, opacity=0.1]
table {%
0 1
1 0.00563187388009265
2 -0.126506009139091
3 -0.0525850694741347
4 -0.0521159122491117
5 0.00449271104258411
6 -0.0616184044693924
7 -0.0645518442760308
8 -0.0391142566079679
9 -0.0389363020188545
10 0.047117671828911
11 -0.144515272111551
12 -0.0671298756794306
13 0.0898306892739764
};
\addplot [semithick, black, opacity=0.1]
table {%
0 1
1 -0.677560389799782
2 0.62998644445765
3 -0.449456717140913
4 0.286931602030493
5 -0.17486203197534
6 -0.0488007347578849
7 -0.0455367900349733
8 -0.0207013827792487
};
\addplot [semithick, black, opacity=0.1]
table {%
0 1
1 -0.320878546240054
2 0.315327969243749
3 -0.194043519609508
4 0.032653200643744
5 -0.138541771536091
6 -0.0345758918314543
7 -0.0609195425392067
8 -0.0786410643434702
9 -0.0380933940867121
10 0.0177125602990049
};
\addplot [semithick, black, opacity=0.1]
table {%
0 1
1 0.13678207578482
2 0.122900059735813
3 0.00825852456715968
4 -0.140846313748377
5 -0.0947544690129613
6 0.130340924088501
7 -0.158459493176853
8 -0.0864800777736791
9 -0.0797781203188791
10 -0.134740344219298
11 -0.10320616442539
12 -0.100016601500857
};
\addplot [semithick, black, opacity=0.1]
table {%
0 1
1 -0.354627878610759
2 -0.0874177102558316
3 0.270791245951533
4 -0.250182564593619
5 -0.0982998690586512
6 -0.0406898787135239
7 0.0367315633853014
8 -0.0165085446975815
9 0.0583031137307655
10 -0.0180994771376331
};
\addplot [semithick, black, opacity=0.1]
table {%
0 1
1 -0.0766560178627362
2 -0.0553720022430132
3 -0.0929650321688064
4 0.0104255609077751
5 -0.0671405189161757
6 0.0205306544829581
7 -0.0440646332290054
8 -0.0220511313237495
9 -0.0824920642448408
10 0.0655856500992564
11 -0.0839209002184759
12 -0.0115339844478728
13 -0.0292560400858476
14 0.0625836755003112
15 0.0950245662889029
16 -0.0209425068912905
17 -0.122216968460109
18 -0.0266371675774378
19 -0.0239809591188683
20 0.0403553490380248
};
\addplot [semithick, black, opacity=0.1]
table {%
0 1
1 0.249354465023295
2 -0.263972147243959
3 0.0367680997364206
4 0.0393685213580965
5 -0.152339129087945
6 -0.226905140024152
7 -0.233734764778332
8 -0.0409677424164808
9 0.0924278374330576
};
\addplot [semithick, black, opacity=0.1]
table {%
0 1
1 -0.298005257242238
2 -0.201994742757762
};
\addplot [semithick, black, opacity=0.1]
table {%
0 1
1 -0.23417452778923
2 -0.115749278168348
3 0.0712489598604552
4 -0.239559541078785
5 0.167182285911689
6 -0.250984299380489
7 -0.02495679896569
8 0.126955109986891
9 0.012851142595355
10 0.101949601447418
11 -0.171551729285511
12 -0.0602833723278637
13 0.11707244719411
};
\addplot [semithick, black, opacity=0.1]
table {%
0 1
1 -0.00470723047561559
2 0.100643154125259
3 -0.266272992821481
4 -0.136022725975646
5 0.152741158088957
6 0.096507539571637
7 0.058261862794156
8 -0.252386610594649
9 -0.117631126097684
10 -0.14144012087625
11 0.010307092261316
};
\addplot [semithick, black, opacity=0.1]
table {%
0 1
1 -0.0429807046463063
2 -0.099048883776086
3 0.0747435727033127
4 -0.154876665370154
5 0.165628925805005
6 -0.0965939245400857
7 -0.0492371444817506
8 -0.0403239340412922
9 -0.0933166264197815
10 0.0598475263223753
11 -0.155569744746687
12 -0.0682723968085494
};
\addplot [semithick, black, opacity=0.1]
table {%
0 1
1 -0.0774799408971063
2 -0.136805313408447
3 0.0923085825800015
4 -0.528094229133174
5 -0.00798176302583203
6 0.0767955911084159
7 -0.0218244637806317
8 0.0976163473283479
9 0.00546518922842608
};
\addplot [semithick, black, opacity=0.1]
table {%
0 1
1 0.403454086586857
2 0.287322659699051
3 0.18075875009724
4 -0.0811689071289936
5 -0.139743587631415
6 -0.272030205909067
7 -0.273908827533323
8 -0.331087518310154
9 -0.273596449870193
};
\addplot [semithick, black, opacity=0.1]
table {%
0 1
1 0.239040608177384
2 -0.296334933512625
3 -0.15410967985084
4 -0.00932870857268253
5 0.111560704279742
6 -0.0407741257648877
7 -0.313967100773502
8 -0.090909342372236
9 0.054822578389647
};
\addplot [semithick, black, opacity=0.1]
table {%
0 1
1 0.44047014708685
2 -0.0830152116288717
3 -0.1527079529423
4 -0.056264884446929
5 -0.120084434668734
6 -0.199514745482318
7 -0.239712869002468
8 -0.0891700489152293
};
\addplot [semithick, black, opacity=0.1]
table {%
0 1
1 0.29819311242213
2 0.00276407225638793
3 -0.0803233447534609
4 -0.0654165779126209
5 -0.357799644131245
6 -0.262847129670275
7 -0.0345704882109168
};
\addplot [semithick, black, opacity=0.1]
table {%
0 1
1 -0.369792584728383
2 -0.0227492575349005
3 0.213686167015368
4 -0.57184874796933
5 0.233623061372261
6 -0.0371652242380743
7 -0.0967357655205401
8 0.1850958294178
9 0.000672005158349685
10 0.0088399729440834
11 -0.00429981110001373
12 -0.00760455405074169
13 -0.00970894952259966
14 -0.0220121412432783
};
\addplot [semithick, black, opacity=0.1]
table {%
0 1
1 -0.212013355919389
2 0.230164181147806
3 0.00549323123268344
4 0.102359878449969
5 -0.124285674275726
6 -0.153236524493857
7 0.286353869841281
8 -0.0930564563387189
9 0.0372817188889158
10 -0.161692556104702
11 0.030408577861835
12 -0.202136319534969
13 -0.207226294874518
14 0.0948340387025149
15 -0.164810735975229
16 0.0315624213921031
};
\addplot [semithick, black, opacity=0.1]
table {%
0 1
1 0.0811360512038072
2 -0.0828683618965547
3 -0.084136885771201
4 -0.114271763151658
5 0.0975825292819362
6 -0.249729974811268
7 -0.22681152649606
8 0.0790999316409985
};
\addplot [semithick, black, opacity=0.1]
table {%
0 1
1 -0.108645844625615
2 0.00284104095728067
3 -0.124309810896252
4 0.072702570643514
5 0.0253787920623272
6 0.0118360764334846
7 -0.0687889212055193
8 0.00472916450748014
9 -0.120821681476915
10 -0.0810811074289549
11 -0.113840278970831
};
\addplot [semithick, black, opacity=0.1]
table {%
0 1
1 0.194219781984058
2 -0.0204827839989106
3 -0.0601919530452339
4 -0.0350286687516383
5 -0.123769052694196
6 -0.15449517149305
7 -0.175364802190789
8 -0.12488734981024
};
\addplot [semithick, black, opacity=0.1]
table {%
0 1
1 -0.152448823334048
2 -0.347551176665952
};
\addplot [semithick, black, opacity=0.1]
table {%
0 1
1 -0.187776871282614
2 0.0187344224827773
3 -0.201050484047407
4 0.00643349027711698
5 0.048570430383414
6 0.0948544880565761
7 -0.0883180591684584
8 -0.0556667755972173
9 0.0339079436258729
10 -0.0773027384500332
11 -0.092385846280028
};
\end{axis}

\end{tikzpicture}